\documentclass[useAMS,usenatbib]{mn2e}

\usepackage{times}              
\usepackage[fleqn]{amsmath}     
\usepackage{amssymb}            
\usepackage{aas_macros}
\usepackage[usenames]{color}
\usepackage{graphicx}
\usepackage{booktabs}
\usepackage[pdftex,bookmarks=true]{hyperref} 

\definecolor{highlight}{rgb}{0.820,0.196,0.314}     

\newcommand{\vertspace}[1]{\renewcommand{\arraystretch}{#1}}

\newcommand{\masyr}{mas\,yr$^{-1}$}
\newcommand{\kms}{km\,s$^{-1}$}
\newcommand{\dgr}{$^\circ$}

\newcommand{\Msunpccube}{M$_\odot$\,pc$^{-3}$}

\newcommand{\du}{\mathrm{d}}               
\newcommand{\afe}{[\ensuremath{\alpha}/Fe]}
\newcommand{\feh}{[Fe/H]}
\newcommand{\atilt}{\ensuremath{\alpha_\mathrm{tilt}}}

\newcommand{\SEGUE}{\textit{SEGUE}}

\def\hdisk{h_\mathrm{disk}}

\def\hthin{h_\mathrm{thin}}

\def\hthick{h_\mathrm{thick}}

\def\rhodm{\rho_\mathrm{dm}}
\def\rhodisk{\rho_\mathrm{disk}}
\def\rhothin{\rho_\mathrm{thin}}
\def\rhothick{\rho_\mathrm{thick}}

\def\zetatr{\zeta_\mathrm{tr}}

\def\Rsun{R_\odot}

\def\mvRz{\overline{v_\mathrm{R}v_\mathrm{z}}}
\def\mvp{\overline{v_\phi}}

\def\vi{\boldsymbol{v}_i}
\def\di{\Delta_i}
\def\bmu{\boldsymbol{\mu}}
\def\bmuj{\bmu_j}
\def\bsig{\boldsymbol{\Sigma}}
\def\bsigj{\bsig_j}

\def\bsigh{\bsig_\mathrm{halo}}
\def\sigrh{\sigma_\mathrm{R,halo}}
\def\sigzh{\sigma_\mathrm{z,halo}}
\def\sigsqrh{\sigma^2_\mathrm{R,halo}}
\def\sigsqzh{\sigma^2_\mathrm{z,halo}}

\def\lih{\mathcal{L}_i^\mathrm{halo}}
\def\lijd{\mathcal{L}_{ij}^\mathrm{disk}}
\def\lij{\mathcal{L}_{ij}}

\title[The tilt of the velocity ellipsoid in the Milky Way disk]{The tilt of the velocity ellipsoid in the Milky Way disk}

\author[A. B\"udenbender et al.]{%
    Alex~B\"udenbender$^1$\thanks{E-mail: buedenbender@mpia.de},
    Glenn~van~de~Ven$^1$, Laura~L.~Watkins$^{1,2}$ \\
    $^1$ Max Planck Institute for Astronomy, K\"onigstuhl 17, 69117 Heidelberg, Germany \\
    $^2$ Space Telescope Science Institute, 3700 San Martin Drive, Baltimore, MD 21218, USA}

\date{Accepted 2015 June. Received 2015 May; in original from 2014 July}

\pagerange{\pageref{firstpage}--\pageref{lastpage}} \pubyear{2015}

\begin{document}

\label{firstpage}

\maketitle

\begin{abstract}
Accurate determination of the local dark matter density is important for understanding the nature and distribution of dark matter in the universe. This requires that the local velocity distribution is characterised correctly. Here, we present a kinematic study of 16,276 SDSS/\SEGUE\ G-type dwarf stars in the solar neighbourhood, with which we determine the shape of the velocity ellipsoid in the meridional plane. We separate our G-dwarf stars based on their \feh{} and \afe{} abundances and infer the local velocity distribution independently for each sub-sample using a maximum-likelihood method that accounts for possible contaminants.

We show by constructing vertical Jeans models that the different sub-samples yield consistent results only when we allow the velocity ellipsoid in the disk to be tilted, demonstrating that the common assumption of decoupled radial and vertical motions in the disk is incorrect. Further, we obtain that the tilt of the velocity ellipsoid is consistent among the different sub-samples. We find that increase in the tilt with height is well described by the relation  $\atilt = (-0.90 \pm 0.04) \arctan(|z|/R_{\sun}) - (0.01 \pm 0.005)$, which is close to alignment with the spherical coordinate system and hence a velocity ellipsoid pointing to the Galactic centre.
We also confirm earlier findings that the sub-samples behave almost isothermally with both radial and vertical velocity dispersion approximately constant with height.

We conclude that the coupling between radial and vertical motion captured in the velocity ellipsoid tilt cannot be ignored when considering dynamical models of the solar neighbourhood. In a subsequent paper, we will develop a new modelling scheme informed by these results and make an improved determination of the local dark matter density.
\end{abstract}


\section{Introduction}
\label{sec:intro}

The concordance cosmological model is based on collisionless dark matter particles, of yet unknown nature, which cannot be detected directly, but which interact through gravity. Various direct detection experiments aim to uncover the nature of these particles, in particular their mass, but, since the signal will depend strongly on their distribution in the Solar neighbourhood, the local dark matter density needs to be measured independently and accurately \citep[e.g.][]{peter2011}. Such a local measurement is also essential to constrain the overall dark matter distribution in the Milky Way as good measurements of the Galactic rotation curve exist but these do not allow the separation of luminous and dark matter due to the so-called disk-halo degeneracy \citep[e.g.][]{dutton2011}.

The traditional approach adopted to measure the local dark matter density is through the vertical force, i.e., the derivative of the gravitational potential away from the Galactic disk plane, inferred from a population of stars with observed vertical number density profile and vertical velocity dispersion profile \citep[e.g.][]{kuijken1989a}. Recent surveys such as the Sloan Extension for Galactic Understanding and Exploration \citep[\SEGUE;][]{yanny2009} make it possible to extract robust vertical density and dispersion profiles even for chemically different subpopulations, providing independent tracers of the same gravitational potential. However, even with many thousands of stars the uncertainties on the dark matter density are still substantial and systematic differences between studies remain even if similar data sets are being used \citep[e.g.][]{zhang2013}.

Most investigations of the local dark matter density to date have used the vertical Jeans equation, which relates the gravitational potential directly to observable vertical profiles without having to specify the phase-space distribution function of the tracers. Unfortunately, the inference of the vertical profiles is often based on taking statistical moments of discrete data within a certain bin, which not only implies loss of information, but is also very sensitive to interlopers. Moreover, the motions of stars in the vertical and radial directions are typically coupled, however often a simple approximation is adopted or the coupling is neglected altogether.

This radial-vertical coupling is reflected in the tilt of the velocity dispersion ellipsoid with respect to the Galactic mid-plane. In turn, this tilt is related to the shape of the gravitational potential, but also depends on the phase-space distribution function. Only in the case of a St\"ackel potential can the shape of the gravitational be directly constrained from the tilt of the velocity ellipsoid \citep[e.g.][]{binney2011}. Even so, aside from measuring the local dark matter density, the velocity ellipsoid is also important for constraining dynamical heating processes \citep[e.g.][]{fuchs1987}, including those that might have led to the thickened Milky Way disk \citep[e.g.][]{liu2012, bovy2012b}. The velocity ellipsoid also enters directly into the asymmetric drift correction of the azimuthal to circular velocity \citep{dehnen1998}. Finally, deviations from axisymmetry due to, for example, spiral structure are encoded in the velocity ellipsoid components \citep{binney2008}.

Previous measurements of the local velocity ellipsoid, and in particular its tilt, have been either over a broad range in height \citep[e.g.][]{siebert2008, carollo2010, casetti-dinescu2011} and/or with very large error bars \citep[e.g.][]{smith2012}. These limitations are partly driven by the limited availability of large samples of stars with reliable photometric and kinematic measurements. For this study, we use a large and well-characterized sample of \SEGUE\ G-type dwarf stars. The method used to extract the velocity moments also plays an important role, so we introduce a discrete likelihood method that explicitly accounts for interlopers and uses a Bayesian inference of the velocity moments.

We describe the G-dwarf sample and kinematic extraction method in \autoref{S:localkin} and construct vertical Jeans models for chemically different sub-samples in \autoref{S:vertjeans}. Even though they are tracers of the same gravitational potential, the inferred value of local dark matter density varies substantially, which we believe mainly to be a consequence of the invalid assumption of decoupled vertical and radial motion. In \autoref{S:tilt}, we indeed confirm that the tilt of the velocity ellipsoid for each sub-sample is non-zero and similarly pointing toward the Galactic centre. In \autoref{S:discconcl}, we discuss how this strongly-improved measurement of the velocity tilt provides important constraints on dynamical models of the Milky Way disk. In the Appendix~\ref{A:nonaxi}, we show that our measurements in the meridional $(R,z)$-plane under the assumption of axisymmetry are affected neither by motion in the azimuthal direction nor by a slight non-zero vertical and radial mean velocities.

Throughout we adopt 8\,kpc for the Sun's distance to the Galactic centre, and 220\,\kms\ for the circular velocity of the local standard of rest (LSR) \citep{kerr1986}. We adopt for the Sun's peculiar velocity relative to the LSR the common values of $(10.00, 5.25, 7.17)$\,\kms\ in the radial, azimuthal and vertical direction, respectively \citep{dehnen1998}.

\section{Local stellar kinematics}
\label{S:localkin}

We briefly introduce the sample of G-type dwarf stars and kinematic extraction algorithms we use to probe the dynamics in a local volume of about 1 kpc in radius around the Sun and from about 0.5 to 2.5 kpc away from the mid-plane.

\subsection{\SEGUE\ G-type dwarf stars}
\label{SS:sample}

The data used in this paper are the same as the \SEGUE\ G-type dwarf data used in \citet{liu2012} to which we refer for further details. In brief, of the wide variety of stars covered by \SEGUE\ \citep{yanny2009}, we focus on G dwarfs as they are abundant and have been targeted for spectroscopy with minimal selection biases. Among possible stellar tracers of the disk dynamics, G dwarfs are the brightest with main-sequence life-times long enough to validate the assumption of dynamical equilibrium. Moreover, their rich metal-line spectrum enables reliable line-of-sight velocities, metallicities \feh, and abundances \afe, with typical uncertainties for S/N$>$15 of 2--5\,\kms, 0.2\,dex, and 0.1\,dex respectively \citep{lee2011}.

\begin{figure}
\begin{center}
    \includegraphics[width=0.95\linewidth]{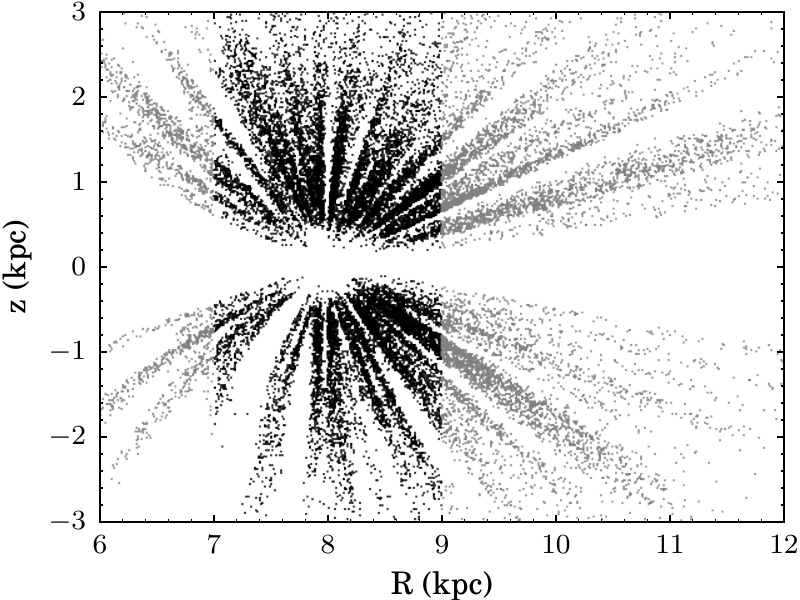}
    \caption{The distribution of the G dwarf sample in the radial and vertical plane. The black points mark the stars inside the region we define as the solar neighbourhood, while we reject the grey points outside the solar neighbourhood.}
    \label{fig:space_dist}
\end{center}
\end{figure}

G dwarfs are selected in SEGUE as stars with $r$-band magnitude $14.0 < r_0 < 20.2$ and colour $0.48 < (g-r) < 0.55$. \autoref{fig:space_dist} illustrates the distribution of the SEGUE G dwarf sample in the local volume: black points are those we consider to be within the solar neighbourhood and are thus used for our analysis. We show the distribution of the resulting sample in \feh\ and \afe\ space in the top panel of \autoref{fig:vert_results}.
We augment our kinematic data with proper motions from the USNO-B survey, which are good to 1--5\,\masyr, while distances based on the photometric colour-metallicity-absolute-magnitude relation of \citet{ivezic2008} have relative errors of $\sim10$\%.

The line-of-sight velocities and proper motions of the stars are transformed into the three velocity components along cylindrical coordinates, namely radial velocity $v_R$, azimuthal or rotational velocity $v_\phi$, and vertical velocity $v_z$. Taking into account the errors in line-of-sight velocities, proper motions and distances, the resulting uncertainties in the velocity components in cylindrical coordinates are on average 10\,\kms. At the furthest distances of $\sim$3\,kpc, the velocity error can increase to 40\,\kms, but no biases are introduced as the velocity error remains smaller than the intrinsic velocity dispersion of the stars.

We focus our analysis on vertical gradients, so that to avoid biases due to radial gradients we concentrate on the Solar cylinder with stars between 7 and 9 kpc from the Galactic centre. In the end, the sample then consists of a total of 16,276 stars between 0.5 and 3.0 kpc away from the mid-plane.

\subsection{Velocity ellipsoid in the meridional plane}
\label{SS:meridionalplane}

We treat the Milky Way disk as an axisymmetric system in a steady state, so that the potential $\Phi \left( R, z \right)$ and the distribution function do not vary with azimuth $\phi$ or time. From \citet{jeans1915}, we then know that the distribution function depends only on isolating integrals of the motion: energy $E = \frac{1}{2} \left( v_R^2 + v_\phi^2 + v_z^2 \right) + \Phi \left( R, z \right)$, angular momentum $L_z = R v_\phi$, and a third integral $I_3$ whose form is not generally known. However, in the absence of resonances, $I_3$ is invariant under the change $\left( v_R, v_z \right) \to \left( -v_R, -v_z \right)$, from which it follows that the mean velocity is in the azimuthal direction $\left( \overline{v_R} = \overline{v_z} = 0 \right)$ and the velocity ellipsoid is aligned with the rotation direction $\left( \overline{v_R v_\phi} = \overline{v_\phi v_z} = 0 \right)$.

The remaining second velocity moment $\mvRz$ then quantifies the coupling between the radial and vertical motions, and, in combination with the radial and vertical velocity dispersion, $\sigma_R$ and $\sigma_z$ yields the tilt of the velocity ellipsoid. We extract the latter velocity moments from the observed radial and vertical velocities, $v_R$ and $v_z$, but do not need to consider the observed azimuthal velocities $v_\phi$, if the Milky Way disk is axisymmetric locally. In Appendix~\ref{A:nonaxi}, we show that excluding or including the azimuthal velocities yields consistent results for $\sigma_R$, $\sigma_z$ and $\mvRz$. Thus, we exclude the azimuthal velocities from the remainder of the current analysis; this is particularly convenient because it is well known that the distribution in $v_\phi$ is non-Gaussian.

The distribution in $v_R$ and $v_z$, on the other hand, is well described by a bi-variate Gaussian. However, $\overline{v_R}$  and $\overline{v_z}$ are observed to be mildly non-zero especially closer to the mid-plane \citep{williams2013}, in line with deviations from axisymmetry due to spiral structures \citep{faure2014}, Even so, in Appendix~\ref{A:nonaxi}, we show that, at the heights $0.5 < |z|/\mathrm{kpc} < 2.5$ probed by the G dwarfs, the deviations are so small that they do not affect the inferred second velocity moments. So to decrease the statistical uncertainty on particular $\mvRz$ and, hence, on the subsequent tilt angle measurement, we set $\overline{v_R} = \overline{v_z} = 0$ for the remainder of the paper.

The only non-zero velocity moments are, thus, second moments $\sigma_R$, $\sigma_z$, $\mvRz$. To determine these velocity moments for a subset of stars (typically selected, in this paper, to have similar heights, metallicities and $\alpha$-element abundances), we use a maximum likelihood approach, which we discuss below.

\subsection{Extracting velocity moments}
\label{SS:likelihoods}

Consider a dataset of $N$ stars where the $i$th star has velocity vector $\vi$ and uncertainty matrix $\di$. Now suppose that the velocity distribution in the disk may be modelled as a multivariate Gaussian $j$ of rank $n$ with mean $\bmuj$ and variance $\bsigj$. We wish to know what is the likelihood that star $i$ came from the disk distribution predicted by Gaussian $j$, which can be written as
\begin{align}
    \lijd & = \mathcal{L} \left( \vi \left| \bmuj, \bsigj, \di \right.
        \right) \nonumber \\
    & = \frac{1}{ \left( 2 \pi \right)^{\tfrac{n}{2}} \left| \bsigj'
        \right| ^{\tfrac{1}{2}} } \exp\left( - \frac{1}{2} \left( \vi - \bmuj
        \right) ^T \bsigj'^{-1} \left( \vi - \bmuj \right) \right) .
    \label{eq:lijd}
\end{align}
where $\bsigj' = \bsigj + \di^2$ results from the convolution of the intrinsic variance of the Gaussian and the observed uncertainties. Here, $\bmuj$ and $\bsigj$ are unknown parameters that we wish to determine.

\begin{figure*}
\begin{center}
    \includegraphics[width=0.48\linewidth]{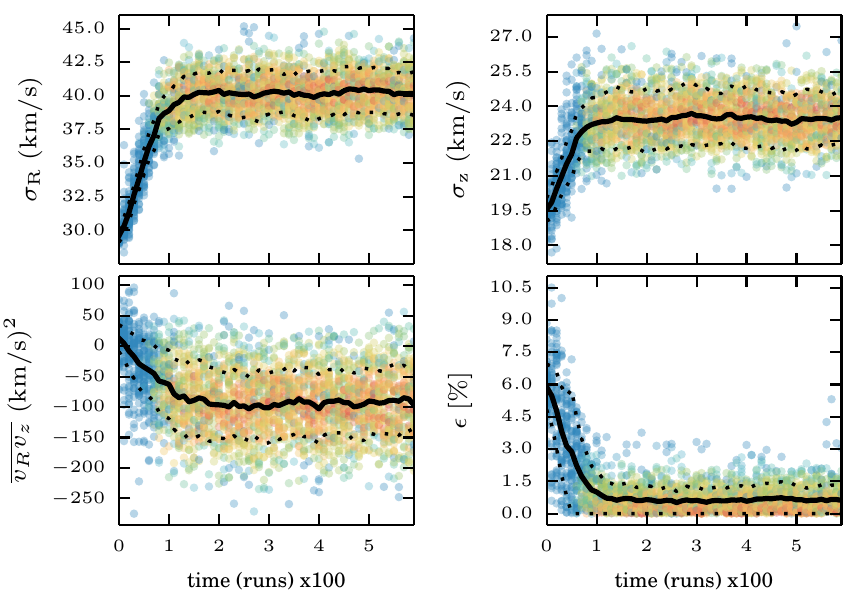}
    \includegraphics[width=0.48\linewidth]{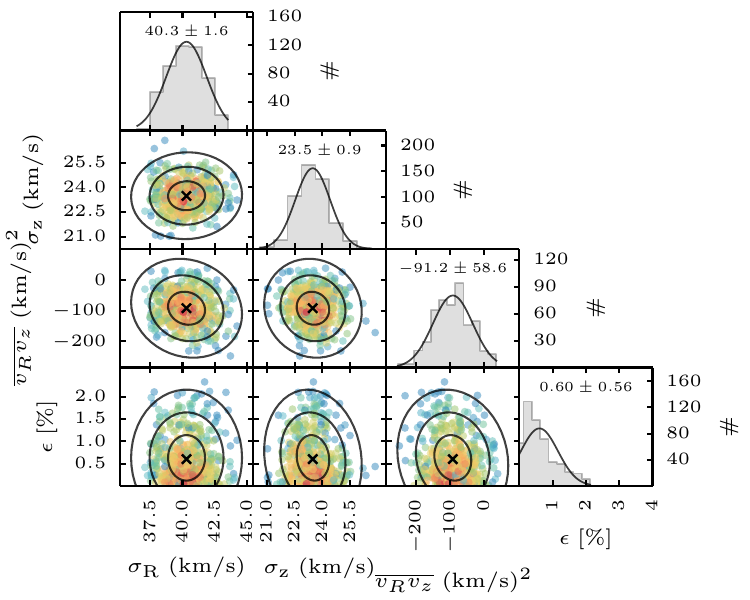}
    \caption{
        \emph{Left:} Parameter evolution in a typical MCMC run. The points show the values visited by the walkers at each step and are coloured by likelihood from red (high) to blue (low). The solid lines show the means at each step and the dotted lines show the dispersions. All parameters converge quickly and tightly.
        \emph{Right:} Post-burn parameter distributions from a typical MCMC run. The scatter plots show the projected two-dimensional distributions of the parameters, with the points coloured by likelihood (red high and blue low). The crosses indicate mean values and the ellipses encompass the 1--3$\sigma$ regions. The histograms show the projected one-dimensional parameter distributions with lines representing gaussians with the same mean and standard deviation. We also give the one-dimensional mean and uncertainty for each of the parameters. We do not see significant correlations between the parameters.
    }
    \label{fig:mcmc}
\end{center}
\end{figure*}

Our dataset is also contaminated by Milky Way halo stars, which we assume to have a Gaussian velocity distribution with a mean of zero and variance $\bsigh$. We also need to consider the likelihood of observing star $i$ given the halo population, which we write as
\begin{align}
    \lih & = \mathcal{L} \left( \vi \left| \bsigh, \di \right.
        \right) \nonumber \\
    & = \frac{1}{ \left( 2 \pi \right)^{\tfrac{n}{2}} \left| \bsigh'
        \right| ^{\tfrac{1}{2}} } \exp\left( - \frac{1}{2} \vi^T \bsigh'^{-1}
        \vi \right) .
    \label{eq:lih}
\end{align}
where $\bsigh' = \bsigh + \di^2$ results from the convolution of the variance of the halo distribution and the observed uncertainties.

We adopt a canonical single halo model with dispersions $\sigrh = 157 \pm 10$~\kms and $\sigzh = 75 \pm 8$~\kms \citep{schoenrich2011}, where $\sigsqrh$ and $\sigsqzh$ are the diagonal elements of $\bsigh$ while the cross term is assumed to be zero. We show in \autoref{S:discconcl} that neither a dual-halo contamination model \citep[e.g.][]{carollo2007, carollo2010} nor the presence of a metal-weak tail to the thick disk \citep[e.g.][]{chiba2000} effects our results.

If we assume that a (small) fraction $\epsilon_j$ of the stars are halo stars -- and so fraction $\left( 1-\epsilon_j \right)$ are disk stars -- then the total likelihood of star $i$ is given by
\begin{equation}
    \lij = \left( 1 - \epsilon_j \right) \lijd + \epsilon_j \lih
\end{equation}
The halo fraction $\epsilon_j$ will be another free parameter in our models. The total likelihood of model $j$ is the product of the model likelihoods for each star
\begin{equation}
    \mathcal{L}_j = \prod_{i=1}^{N} \lij .
\end{equation}
The best model is that which maximises $\mathcal{L}_j$.

In general, our free parameters are $\bmuj$, $\bsigj$ and $\epsilon_j$. However, as we discussed in \autoref{SS:meridionalplane}, we can assume that all components of $\bmuj$ and a number of elements of $\bsigj$ are zero. So, in practice, we have only four free parameters for each model $j$: $\sigma_R$, $\sigma_z$, $\mvRz$ and $\epsilon$. In order to efficiently sample our parameter space as we search for the best model, we use a Markov Chain Monte Carlo (MCMC) analysis; we use the \textsc{emcee} package developed by \citet{foremanmackey2013}, which is an implementation of the affine-invariant MCMC ensemble sampler by \citet{goodman2010}. Our MCMC chains use 100 walkers and run for 600 steps. We consider the first 500 steps as the burn-in phase that finds the region of parameter space where the likelihood is highest. The final 100 steps then constitute the post-burn phase that explores the high-likelihood region.

\autoref{fig:mcmc} illustrates the output from an MCMC run on a typical subset of our kinematic data (around 500 stars). The left-hand panels show the evolution and eventual convergence of the MCMC chain. The coloured points show the values sampled by the walkers at each step with the colours representing the likelihood of the model (red high and blue low). The solid lines show the means of the walker values and the dotted lines show the 1$\sigma$ dispersions of the walker values. All of the parameters converge tightly. The right-hand panels show the post-burn parameter distributions. The scatter plots show the two-dimensional distributions of the parameters, again with points coloured according to their likelihoods (red high and blue low). The ellipses show the $1\sigma$, $2\sigma$ and $3\sigma$ regions of the covariance matrix for the post-burn parameter distribution, projected into each 2D plane. The crosses mark the means of the parameter distributions. The histograms show the one-dimensional distributions of the parameters; the solid black lines represent Gaussians with the same mean and standard deviation. The histogram panels also give the one-dimensional mean and uncertainty for each of the parameters.

\section{Vertical Jeans model}
\label{S:vertjeans}

We use dynamical models to link observable quantities (such as stellar number density $\nu$ and velocity dispersion $\sigma$) with quantities that we wish to know but are not able to measure directly (such as mass density $\rho$ and potential $\Phi$).

Stellar sub-samples with different origins as reflected in their different ages and/or chemical properties, will have different spatial distributions ($\nu$) and different kinematics ($\sigma$). Nevertheless, they feel the same underlying total mass density that gives rise to the same underlying gravitational potential. So, in theory, if we use the observed kinematics of a number of sub-samples independently to find the best-fit density distribution in the solar neighbourhood, all sub-samples should return the same answer. However, in practice, we will only obtain consistent results from the different sub-samples if the assumptions we make in the modelling are correct.

Our goal here is to assess the validity of the assumption that the radial and vertical motions of stars in the Milky Way disk are decoupled. As such, we first select two sub-samples of G-dwarf stars based on their \feh\ metallicities and \afe\ abundances. Then we model the local mass density independently for the two sub-samples, assuming that the vertical and radial motions are decoupled, and test the agreement of the two best-fit models.

\subsection{Gravitational potential}
\label{SS:vert_potential}

The total mass density in the solar neighbourhood has contributions from both luminous and dark matter. \citet{juric2008} calculated photometric parallax distances for $\sim$48 million stars selected from the SDSS to determine the 3-dimensional number density distribution of the Milky Way. Using a sub-sample of nearby M-dwarfs, they found that the solar neighbourhood mass density is best described as two exponential disks: a thin disk with density $\rhothin$ and a thick disk with density $\rhothick$, where the fraction of thick disk stars relative to thin disk stars in the plane at the solar radius $\Rsun$ is $f = 0.12$. The thin disk component has a vertical scale height $\hthin = 300$~pc and the thick disk component has a vertical scale height $\hthick = 900$~pc. We adopt this as the stellar density distribution for our analysis\footnote{Note, we assume that all of our stars are at the solar radius, so we neglect any radial variations in disk density.}. Dark matter also makes a contribution $\rhodm$ to the local density distribution; as the radial extent of our data is small and the vertical extent is less than 2~kpc, we can assume that this is constant throughout the region of interest. Thus the total mass density in the solar neighbourhood is given by
\begin{equation}
    \rho_\odot \left( z \right) = \rho \left( \Rsun, z \right) = \rhothin
        \left( \Rsun, z \right) + \rhothick \left( \Rsun, z \right) + \rhodm
\end{equation}
where the thin and thick disk densities are given by
\begin{equation}
    \rhodisk \left( \Rsun, z \right) = \rhodisk \left( \Rsun, 0 \right)
        \exp \left( - \frac{z}{\hdisk} \right)
\end{equation}
and where $\rhodisk \left( \Rsun, 0 \right)$ is the density of the disk component in the plane at the solar radius.

Recalling that we know the local normalisation fraction $f$ of the thick disk relative to the thin disk in the plane
\begin{equation}
    f = \frac{\rhothick \left( \Rsun, 0 \right)}{\rhothin \left( \Rsun, 0 \right)},
\end{equation}
then
\begin{equation}
    \rho_\odot \left( z \right) = \rho_0
        \left[ \exp \left( - \frac{z}{\hthin} \right)
        + f \exp \left( - \frac{z}{\hthick} \right) \right]
        + \rhodm
        \label{eq:localdens}
\end{equation}
where $\rho_0 = \rhothin \left( \Rsun, 0 \right)$.

The potential generated by this density distribution can then be calculated via Poisson's equation
\begin{equation}
    \nabla^2 \Phi = 4 \pi G \rho_\odot.
    \label{eq:poisson}
\end{equation}
We are not able to measure $\Phi$ directly. Instead, we use dynamical models to predict the observable quantities generated in a given potential, then we compare the values we actually observe with those we predict. For our present study, we use the Jeans equations to carry out the dynamical modelling.

Under the assumption of axial symmetry, the vertical first moment Jeans equation in cylindrical polars is
\begin{equation}
    \frac{1}{R} \frac{\partial}{\partial R} \left( R \, \nu \, \overline{v_R
        v_z} \right) + \frac{\partial}{\partial z} \left( \nu \,
        \sigma_z^2 \right) + \nu \frac{\partial \Phi}{\partial z} = 0 .
\end{equation}

If we assume that the velocity ellipsoid is aligned with the cylindrical coordinate system (and hence that radial and vertical motions can be decoupled) then $\overline{v_R v_z} = 0$. Our sample is restricted to the solar neighbourhood and we assume that all stars are at the solar radius $\Rsun$. Hence, the vertical Jeans equation becomes
\begin{equation}
    \frac{\du}{\du z} \left( \nu \, \sigma_z^2 \right)
        + \nu \frac{\du \Phi}{\du z} = 0 .
    \label{eq:vertJeans}
\end{equation}
As we can see, we are actually interested in the first derivative of the potential here, which we calculate from equations \ref{eq:localdens} and \ref{eq:poisson} as
\begin{align}
    \frac{\du \Phi}{\du z} \left( z \right) = & {} 4 \pi G \rho_0 \left\{
        \hthin \left[ 1 - \exp \left( - \frac{z}{\hthin} \right) \right]
        \right. \nonumber \\
    & \left. + f \hthick \left[ 1 - \exp \left( - \frac{z}{\hthick} \right)
        \right] \right\} + 4 \pi G \rhodm z .
    \label{eq:dphidz}
\end{align}

Finally, we need the tracer number density $\nu$ and the vertical velocity dispersion $\sigma_z$; both of which we are able to calculate from observations. Note that different stellar populations may have different number density profiles and different dispersion profiles due to differences in their origins, however they all orbit within the same potential. This point is key to our analysis. By applying these models to multiple stellar sub-samples independently, we can obtain multiple independent estimates for the potential of the system. If the assumptions we have made in the modelling are correct -- principally that the radial and vertical motions may be decoupled -- and equation \ref{eq:vertJeans} is a good representation of reality, then the estimates of the potential should be in good agreement. However, if the potential estimates we recover do not agree, then we can conclude that our assumptions were incorrect.

\begin{figure*}
\begin{center}
    \includegraphics[width=0.45\linewidth]{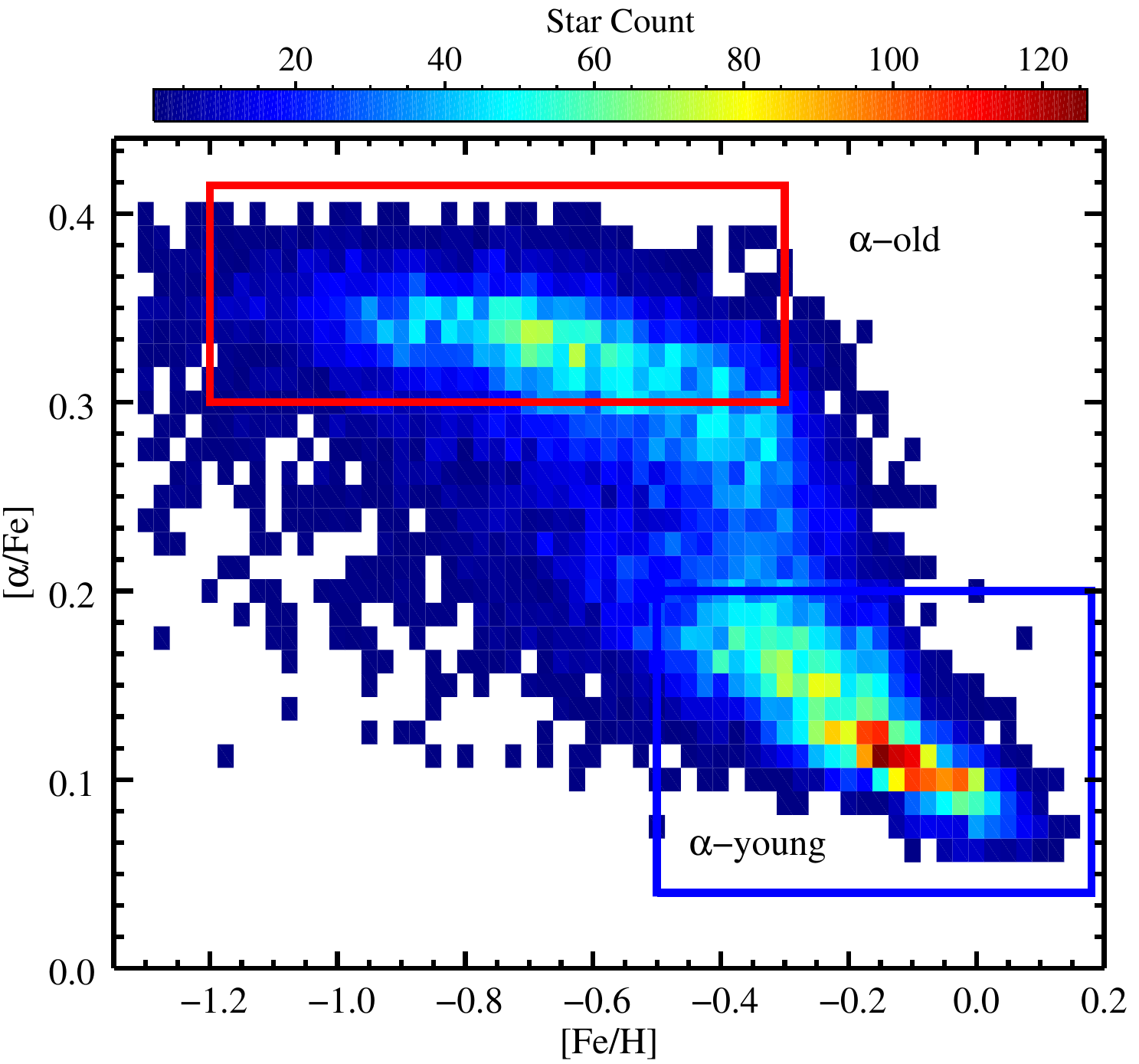} \\
    \includegraphics[width=0.45\linewidth]{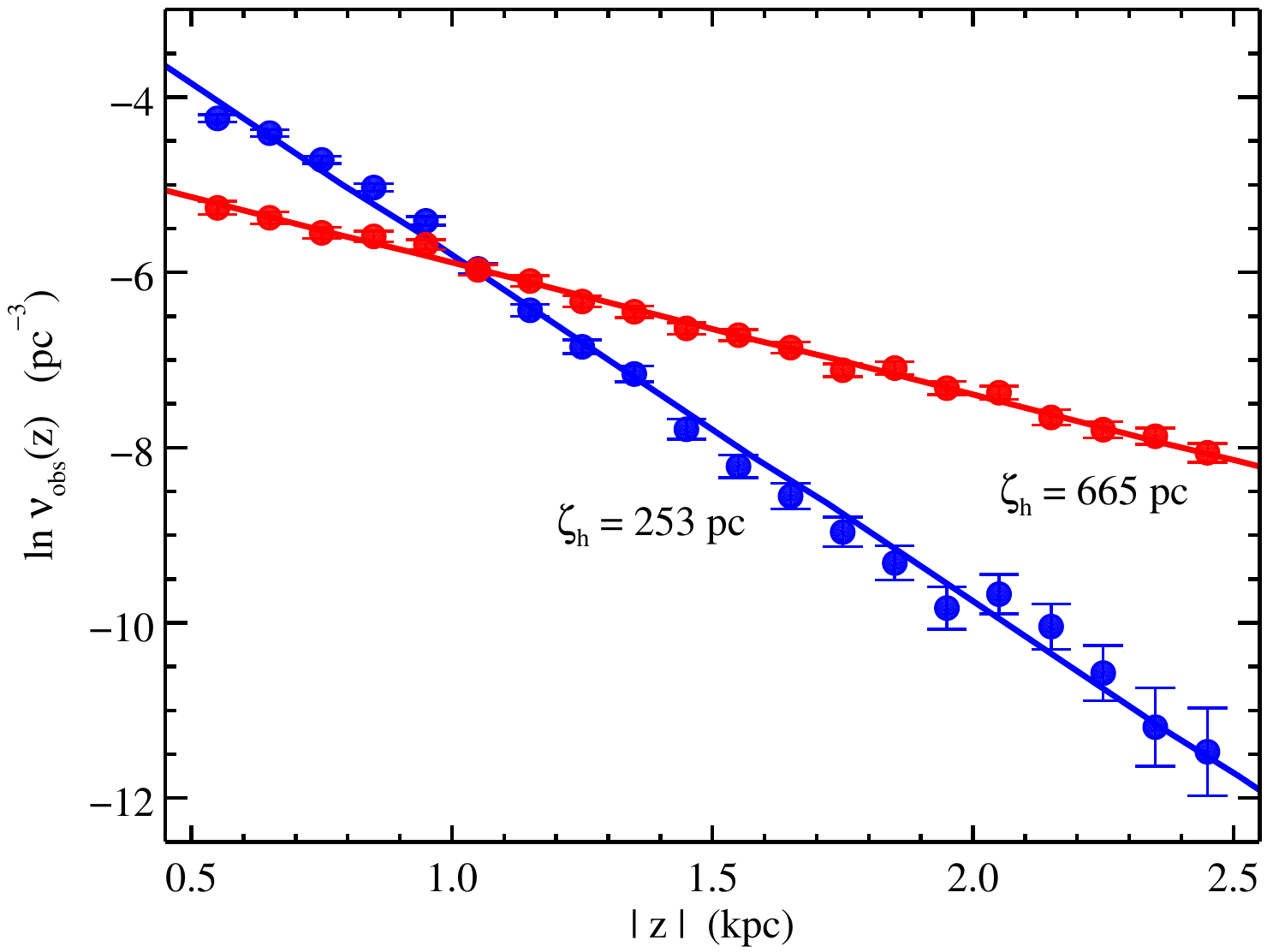}
    \includegraphics[width=0.45\linewidth]{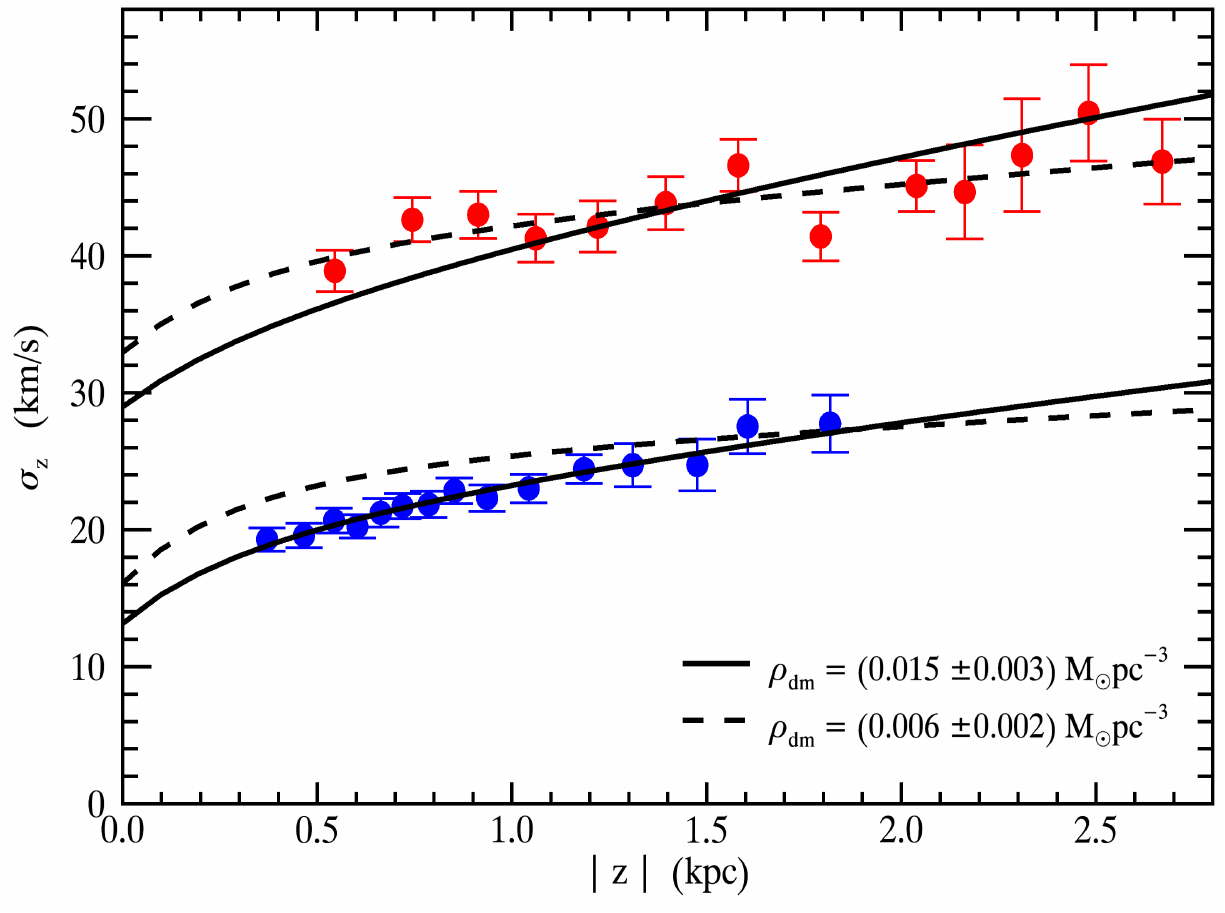}
    \caption{\textit{Top:} \afe\ abundances and \feh\ metallicities of 16,276 SDSS/\SEGUE\ G-dwarf stars, binned in 0.025\,dex by 0.0125\,dex pixels. The pixel colours represent the number counts, as shown by the colour bar. The selection boxes used to extract the two sub-samples we use in this section are shown as red and blue rectangles. $\alpha$-element and iron abundances can be used as a proxy for age; the sub-sample with high \afe\ and low \feh\ we call the $\alpha$-young sub-sample and the sub-sample with low \afe\ and high \feh\ we call the $\alpha$-old sub-sample. 
    \textit{Bottom left:} The selection-function-corrected number density profiles of the $\alpha$-old sub-sample (red) and $\alpha$-young sub-sample (blue). The solid lines are exponential fits with scale heights $\zeta_h$ indicated.
    \textit{Bottom right:} Vertical velocity dispersion as a function of height. The $\alpha$-old sub-sample (red) is best fit by a model with negligible dark matter (upper dashed line) and $\alpha$-young sub-sample (blue) is best fit by a model including dark matter (lower solid line). To aid visual comparison of the models, the upper solid line (lower dashed line) shows the best-fitting $\alpha$-young ($\alpha$-old) density model using the $\alpha$-old ($\alpha$-young) tracer density. As the sub-samples orbit in the same underlying potential, they should make consistent predictions about the local dark matter density. These models assume that the radial and vertical motions can be decoupled; the discrepancy in the fits indicates that this assumption is incorrect.
    }
    \label{fig:vert_results}
\end{center}
\end{figure*}

\subsection{Tracer populations}
\label{SS:vert_tracers}

The top panel of \autoref{fig:vert_results} shows the \afe\ abundances and \feh\ metallicities of the stars in our sample. The stars have been binned into pixels of 0.025~dex by 0.0125~dex and the pixels coloured according to the number of stars in that pixel as shown by the colour bar. $\alpha$-element and iron abundances are particularly useful as they can be used as a proxy for age: stars towards the top-left of parameter space as plotted are older, in general, than the stars towards the bottom-right \citep{loebman2011}. In our sample, there are two clear overdensities: the first occurs at high \afe{} and low \feh{}, representing an older population; the second occurs at high \feh{} and low \afe{}, representing a younger population.

We select two sub-samples centred on these overdensities: the $\alpha$-old sample contains stars with $0.3 <$ \afe\ and $-1.2 <$ \feh\ $< -0.3$; the  $\alpha$-young sample contains stars with \afe\ $< 0.2$ and \feh\ $> -0.5$. These selection boxes are shown in the top panel of \autoref{fig:vert_results}, with the $\alpha$-old selection shown in red and the $\alpha$-young selection shown in blue. For consistency, these colours will be used in all plots comparing results from these two sub-samples.

We assume that the number density $\nu$ of stars in each tracer sub-sample follows an exponential profile such that
\begin{equation}
    \nu \left( z \right) = \nu_0 \exp \left( - \frac{z}{\zetatr} \right)
    \label{eq:tracernum}
\end{equation}
where $\nu_0$ is the number density in the Galactic plane and the $\zetatr$ is the scale height of the tracer sample. To determine the scale-height parameters for each sub-sample, we calculate the number density of stars in a series of height bins and find the best-fitting exponential profile. The number density is highly sensitive to the selection function for SEGUE; to correct for this, we adopt the approach described in Section 3.1.2 of \citet{zhang2013}. The bottom left panel of \autoref{fig:vert_results} shows the logarithm of the corrected number density as a function of vertical distance from the plane for the two sub-samples. The $\alpha$-old sub-sample is shown in red and the $\alpha$-young sub-sample is shown in blue. The data are shown as symbols and the best-fit profiles are shown as solid lines. We find a best-fitting scale height of $\zetatr = 253 \pm 6$~pc for the $\alpha$-young sub-sample and $\zetatr = 665 \pm 11$~pc for the $\alpha$-old sub-sample.

\subsection{Vertical velocity dispersion}
\label{SS:vert_dispersions}

Now that we have a functional form for the tracer density (equation (\ref{eq:tracernum})), we can substitute this and the first derivative of the potential from equation (\ref{eq:dphidz}) into the vertical Jeans equation (\ref{eq:vertJeans}). Rearranging and performing the necessary integration, we obtain a prediction for the vertical velocity dispersion as a function of height
\begin{align}
    \sigma^2_z(z) = & {} 4 \pi G \rho_0 \zetatr \left\{ \hthin
        \left[ 1 - \frac{\hthin}{\hthin + \zetatr} \exp \left( - \frac{z}{\hthin} \right) \right] \right. \notag \\
    & \left. + f \hthick \left[ 1 - \frac{\hthick}{\hthick + \zetatr} \exp \left( - \frac{z}{\hthick} \right) \right] \right\} \notag \\
    & + 4 \pi G \rhodm \zetatr \left( z + \zetatr \right) .
    \label{eq:solved_vertJeans}
\end{align}
There are two free parameters in this expression: the local thin disk density in the plane $\rho_0$ and the local dark matter density $\rhodm$.

To obtain vertical velocity dispersion profiles for our data, we bin the stars in height and use the maximum likelihood method described in \autoref{SS:likelihoods} to calculate the velocity dispersion in each bin. We use 10 bins, with the bin boundaries selected so that each bin contains an equal number of stars. This is done independently for each of our sub-samples. Note that, although we are only interested here in the vertical velocity dispersion $\sigma_z$, our maximum likelihood analysis uses all of the data available and fits for the radial dispersion, the covariance and the background fraction\footnote{The estimated background fraction varies little from bin to bin and never exceeds 5\%.} as well. The bottom-right panel of \autoref{fig:vert_results} shows the vertical velocity dispersion profiles for our two sub-samples; the $\alpha$-young sample is shown in blue and the $\alpha$-old sample is shown in red.

We wish to compare the model predictions against our data and determine which ($\rho_0$, $\rhodm$) values provide a best fit to the observed profile for each sub-sample. We do this using a non-linear least squares (NNLS) fit.

We find that the $\alpha$-old sample is best described by a model with central disk density $\rho_0 = 0.12 \pm 0.011$~\Msunpccube and local dark matter density $\rhodm = 0.0024 \pm 0.0021$~\Msunpccube. This model is shown as dashed lines in the bottom-left panel of \autoref{fig:vert_results}. The upper dashed line is plotted using the value of $\zetatr$ found to best fit the $\alpha$-old sample; as expected, this is is an excellent fit to the $\alpha$-old dispersion profile. In order to show the ability of this model to reproduce the $\alpha$-young profile, the lower dashed line is plotted using the $\alpha$-young $\zetatr$. This is a very poor fit to our $\alpha$-young sample.

We find that the $\alpha$-young sample is best described by a model with central disk density $\rho_0 = 0.06 \pm 0.011$~\Msunpccube and local dark matter density $\rhodm = 0.014 \pm 0.004$~\Msunpccube. This model is shown as solid lines in the bottom-left panel of \autoref{fig:vert_results}. Again, we plot this model using both the $\alpha$-old $\zetatr$ (upper solid line) and the $\alpha$-young $\zetatr$ (lower solid line). This model is an excellent approximation to the $\alpha$-young sample, but fails to reproduce the $\alpha$-old sample.

As we previously discussed, the $\alpha$-old and $\alpha$-young sub-samples feel the same underlying dark-matter density. If our modelling approach is correct and the radial and vertical motions can be decoupled, then the best-fit models determined from the two sub-samples should be consistent. However, we find that the dark matter densities estimated by the two sub-samples are inconsistent: the $\alpha$-young sub-sample favours a model with small but non-negligible local dark matter density, whereas the $\alpha$-old sub-sample favours a model that is consistent with no local dark matter. From this we conclude that our assumption was incorrect and, thus, that the radial and vertical motions cannot be treated independently. This, in turn, implies that the velocity ellipsoid is tilted.

\section{Velocity ellipsoid tilt}
\label{S:tilt}

The coupling between the radial and vertical motions is characterised by the tilt angle $\atilt$ of the velocity ellipsoid defined as
\begin{equation}
    \tan(2 \atilt) = \frac{2 \, \overline{v_R v_z}}{\sigma_R^2 - \sigma_z^2}.
    \label{eq:tiltangledef}
\end{equation}
We expect $\sigma_R$ and $\sigma_z$ to be larger for an older population of stars as a result of internal and external dynamical heating mechanisms over time \citep[e.g.][]{carlberg1985}, as well as due to the possibility that the earliest stars were born dynamical hotter from a more turbulent disk at higher redshift \citep[e.g.][]{foersterschreiber2009}. However, the tilt angle can still be and remain the same for different populations, and, actually, if the (local) potential is of separable St\"ackel form, has to be same. Hence, we now investigate the velocity ellipsoid for different sub-samples independently and find that, within the measurement uncertainties, the title angle is the same. We then combine the sub-samples to arrive at a measurement of the tilt angle, which we show to be consistent but significantly more precise than previous determinations.

\subsection{Velocity ellipsoid of different sub-samples}
\label{SS:velellsubpop}

\begin{table*}
    \caption{Measured velocity ellipsoid components as function of height above the Galactic plane for chemically different sub-samples from \autoref{fig:velellsubpop}. The seven sub-samples are ordered in this table top down from metal-rich and $\alpha$-poor to metal-poor and $\alpha$-rich. The stars within each sub-sample are subdivided in different height ranges (with mean and spread indicated) after which the velocity ellipsoid components in the meridional plane are computed using the likelihood approach described in \autoref{SS:likelihoods}; the mean and standard-deviation of the MCMC post-burn parameter distribution are given. The tilt angle $\atilt$ follows from combing the velocity ellipsoid components as in equation~(\ref{eq:tiltangledef}). }
    \label{tab:velellsubpop}
    \vertspace{1.1}
    \begin{tabular}{@{}cccccccc@{}}
	\toprule
        $\overline{\rm [Fe/H]}$ & $\overline{\rm [\alpha/Fe]}$ & z & $\sigma_{\rm R}$ & $\sigma_{\rm z}$ & $\langle v_{\rm R} v_{\rm z} \rangle$ & $\alpha_{\rm tilt}$ & $\epsilon$ \\
	(dex) & (dex) & (pc) & (\kms) & (\kms) & (\kms) & (deg) & (\%) \\
  \midrule
-0.07 & 0.11 &   449 $\pm$ 124 &   33.5 $\pm$ 1.3 &   19.0 $\pm$ 0.8 &   -40 $\pm$  40 &   -3.0 $\pm$ 2.9 &   1.7 $\pm$ 0.8 \\
      &      &   565 $\pm$  89 &   34.5 $\pm$ 1.4 &   18.9 $\pm$ 0.9 &   -61 $\pm$  42 &   -4.1 $\pm$ 2.8 &   1.7 $\pm$ 0.8 \\
      &      &   667 $\pm$  83 &   37.1 $\pm$ 1.5 &   19.6 $\pm$ 0.9 &   -84 $\pm$  45 &   -4.8 $\pm$ 2.6 &   1.2 $\pm$ 0.8 \\
      &      &   766 $\pm$  97 &   37.8 $\pm$ 1.5 &   20.3 $\pm$ 0.9 &   -92 $\pm$  46 &   -5.1 $\pm$ 2.5 &   1.3 $\pm$ 0.7 \\
      &      &   966 $\pm$ 357 &   38.1 $\pm$ 1.3 &   21.3 $\pm$ 0.8 &  -110 $\pm$  45 &   -6.3 $\pm$ 2.5 &   1.0 $\pm$ 0.5 \\
\hline
-0.21 & 0.14 &   447 $\pm$ 125 &   41.1 $\pm$ 1.6 &   19.4 $\pm$ 0.9 &   -60 $\pm$  46 &   -2.6 $\pm$ 2.0 &   1.0 $\pm$ 0.7 \\
      &      &   563 $\pm$  83 &   40.7 $\pm$ 1.6 &   20.8 $\pm$ 0.9 &   -10 $\pm$  48 &   -0.5 $\pm$ 2.2 &   0.9 $\pm$ 0.6 \\
      &      &   650 $\pm$  71 &   41.3 $\pm$ 1.6 &   21.1 $\pm$ 0.9 &    25 $\pm$  49 &    1.2 $\pm$ 2.2 &   0.5 $\pm$ 0.4 \\
      &      &   739 $\pm$  73 &   43.7 $\pm$ 1.9 &   21.2 $\pm$ 1.0 &   -87 $\pm$  54 &   -3.4 $\pm$ 2.1 &   1.0 $\pm$ 0.9 \\
      &      &   826 $\pm$  78 &   43.8 $\pm$ 1.7 &   24.7 $\pm$ 1.0 &  -177 $\pm$  61 &   -7.6 $\pm$ 2.5 &   0.5 $\pm$ 0.5 \\
      &      &   928 $\pm$ 104 &   42.9 $\pm$ 1.6 &   23.8 $\pm$ 1.0 &    -7 $\pm$  61 &   -0.3 $\pm$ 2.7 &   0.8 $\pm$ 0.6 \\
      &      &  1082 $\pm$ 158 &   41.9 $\pm$ 1.6 &   24.9 $\pm$ 1.0 &    28 $\pm$  63 &    1.4 $\pm$ 3.2 &   0.5 $\pm$ 0.4 \\
      &      &  1328 $\pm$ 394 &   43.3 $\pm$ 1.9 &   26.6 $\pm$ 1.1 &  -106 $\pm$  68 &   -5.1 $\pm$ 3.2 &   0.4 $\pm$ 0.4 \\
\hline
-0.36 & 0.18 &   499 $\pm$ 205 &   37.0 $\pm$ 1.5 &   22.4 $\pm$ 1.0 &   -47 $\pm$  50 &   -3.1 $\pm$ 3.3 &   1.9 $\pm$ 1.0 \\
      &      &   637 $\pm$ 176 &   39.4 $\pm$ 1.6 &   24.7 $\pm$ 1.0 &  -105 $\pm$  53 &   -6.3 $\pm$ 3.1 &   1.0 $\pm$ 0.7 \\
      &      &   761 $\pm$ 195 &   40.5 $\pm$ 1.6 &   24.4 $\pm$ 1.1 &  -131 $\pm$  57 &   -7.1 $\pm$ 3.0 &   0.9 $\pm$ 0.7 \\
      &      &   893 $\pm$ 264 &   40.5 $\pm$ 1.7 &   23.5 $\pm$ 1.1 &  -105 $\pm$  59 &   -5.5 $\pm$ 3.0 &   0.8 $\pm$ 0.7 \\
      &      &  1188 $\pm$ 640 &   41.2 $\pm$ 1.6 &   24.8 $\pm$ 1.0 &  -153 $\pm$  56 &   -7.9 $\pm$ 2.8 &   3.1 $\pm$ 1.1 \\
\hline
-0.35 & 0.28 &   685 $\pm$ 158 &   49.5 $\pm$ 1.9 &   33.9 $\pm$ 1.3 &   -58 $\pm$  86 &   -2.5 $\pm$ 3.8 &   0.5 $\pm$ 0.5 \\
      &      &   892 $\pm$ 104 &   50.0 $\pm$ 1.9 &   32.7 $\pm$ 1.3 &   -66 $\pm$  87 &   -2.6 $\pm$ 3.4 &   1.1 $\pm$ 0.8 \\
      &      &  1106 $\pm$ 106 &   51.5 $\pm$ 2.1 &   34.0 $\pm$ 1.4 &   -74 $\pm$  97 &   -2.8 $\pm$ 3.7 &   0.8 $\pm$ 0.8 \\
      &      &  1362 $\pm$ 134 &   55.7 $\pm$ 2.4 &   34.9 $\pm$ 1.5 &  -199 $\pm$ 116 &   -6.0 $\pm$ 3.4 &   1.4 $\pm$ 1.2 \\
      &      &  1830 $\pm$ 489 &   54.5 $\pm$ 2.6 &   35.3 $\pm$ 1.4 &  -442 $\pm$ 114 &  -13.5 $\pm$ 3.1 &   1.4 $\pm$ 1.1 \\
\hline
-0.51 & 0.29 &   558 $\pm$ 179 &   40.2 $\pm$ 1.7 &   30.8 $\pm$ 1.3 &  -147 $\pm$  72 &  -11.9 $\pm$ 5.3 &   3.2 $\pm$ 1.5 \\
      &      &   734 $\pm$ 139 &   43.8 $\pm$ 2.0 &   32.8 $\pm$ 1.5 &  -139 $\pm$  88 &   -9.1 $\pm$ 5.6 &   2.8 $\pm$ 1.6 \\
      &      &   898 $\pm$ 136 &   47.4 $\pm$ 2.0 &   36.4 $\pm$ 1.4 &     9 $\pm$  97 &    0.6 $\pm$ 6.0 &   1.0 $\pm$ 0.9 \\
      &      &  1065 $\pm$ 149 &   48.1 $\pm$ 2.3 &   34.7 $\pm$ 1.6 &   -42 $\pm$  99 &   -2.2 $\pm$ 5.1 &   2.6 $\pm$ 1.6 \\
      &      &  1254 $\pm$ 174 &   45.8 $\pm$ 2.5 &   33.0 $\pm$ 1.5 &  -130 $\pm$  99 &   -7.2 $\pm$ 5.4 &   4.0 $\pm$ 1.7 \\
      &      &  1490 $\pm$ 232 &   46.4 $\pm$ 2.5 &   36.4 $\pm$ 1.6 &   -15 $\pm$ 111 &   -1.0 $\pm$ 7.6 &   1.8 $\pm$ 1.4 \\
      &      &  1977 $\pm$ 587 &   53.8 $\pm$ 3.0 &   39.6 $\pm$ 1.7 &  -290 $\pm$ 116 &  -11.8 $\pm$ 4.4 &   3.1 $\pm$ 2.0 \\
\hline
-0.68 & 0.32 &   623 $\pm$ 223 &   55.5 $\pm$ 2.4 &   39.4 $\pm$ 1.6 &  -217 $\pm$ 116 &   -7.9 $\pm$ 4.1 &   1.7 $\pm$ 1.4 \\
      &      &   822 $\pm$ 148 &   53.9 $\pm$ 2.3 &   40.4 $\pm$ 1.6 &  -222 $\pm$ 120 &   -9.6 $\pm$ 4.9 &   2.0 $\pm$ 1.3 \\
      &      &   986 $\pm$ 146 &   54.8 $\pm$ 2.5 &   41.4 $\pm$ 1.8 &  -160 $\pm$ 130 &   -7.0 $\pm$ 5.5 &   3.9 $\pm$ 2.0 \\
      &      &  1169 $\pm$ 156 &   57.0 $\pm$ 2.4 &   39.0 $\pm$ 1.8 &  -251 $\pm$ 127 &   -8.1 $\pm$ 4.0 &   3.1 $\pm$ 1.8 \\
      &      &  1367 $\pm$ 178 &   62.2 $\pm$ 2.6 &   42.9 $\pm$ 1.8 &  -456 $\pm$ 151 &  -12.1 $\pm$ 3.7 &   1.8 $\pm$ 1.5 \\
      &      &  1580 $\pm$ 201 &   61.2 $\pm$ 3.0 &   45.1 $\pm$ 1.9 &  -480 $\pm$ 161 &  -14.6 $\pm$ 4.2 &   2.2 $\pm$ 1.8 \\
      &      &  1857 $\pm$ 256 &   61.9 $\pm$ 2.9 &   43.7 $\pm$ 1.8 &  -221 $\pm$ 149 &   -6.5 $\pm$ 4.2 &   1.5 $\pm$ 1.4 \\
      &      &  2225 $\pm$ 449 &   64.5 $\pm$ 3.2 &   43.8 $\pm$ 1.8 &  -608 $\pm$ 171 &  -14.2 $\pm$ 3.5 &   1.7 $\pm$ 1.5 \\
\hline
-0.89 & 0.34 &   817 $\pm$ 272 &   65.1 $\pm$ 3.2 &   46.3 $\pm$ 1.9 &  -407 $\pm$ 179 &  -10.6 $\pm$ 4.3 &   4.5 $\pm$ 2.6 \\
      &      &  1093 $\pm$ 230 &   66.6 $\pm$ 3.8 &   46.5 $\pm$ 1.9 &  -344 $\pm$ 179 &   -8.5 $\pm$ 4.2 &   4.6 $\pm$ 2.7 \\
      &      &  1378 $\pm$ 242 &   67.1 $\pm$ 4.0 &   48.9 $\pm$ 2.0 &  -468 $\pm$ 202 &  -12.0 $\pm$ 4.7 &   5.9 $\pm$ 3.5 \\
      &      &  1675 $\pm$ 251 &   71.4 $\pm$ 3.3 &   46.0 $\pm$ 1.9 &  -337 $\pm$ 186 &   -6.4 $\pm$ 3.4 &   2.1 $\pm$ 1.9 \\
      &      &  2113 $\pm$ 583 &   74.4 $\pm$ 3.7 &   49.0 $\pm$ 2.0 &  -927 $\pm$  68 &  -15.3 $\pm$ 1.4 &   4.4 $\pm$ 3.4 \\
\bottomrule
    \end{tabular}
\end{table*}

\begin{figure*}
\begin{center}
        \includegraphics[width=0.45\linewidth]{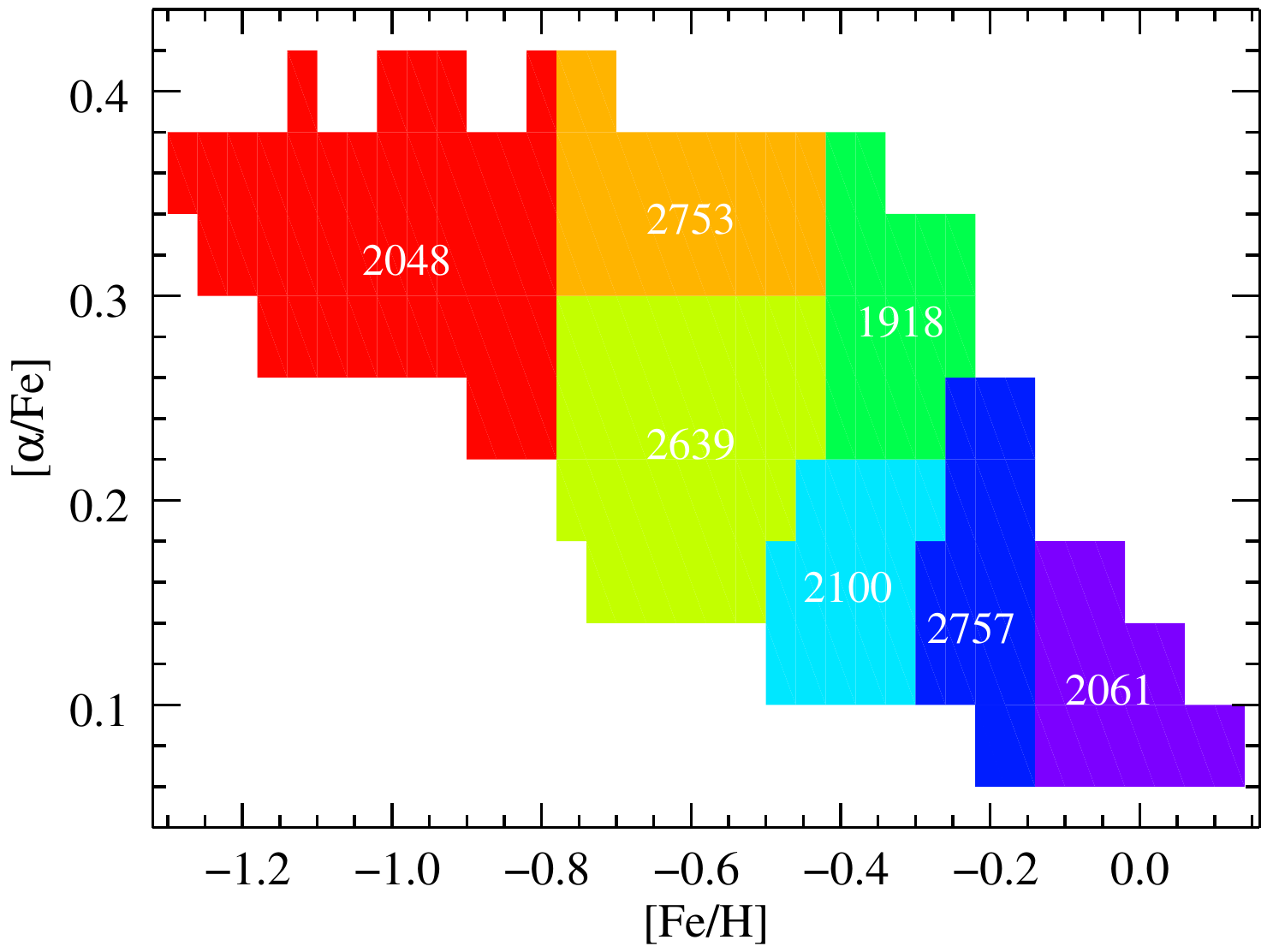}
    \includegraphics[width=0.45\linewidth]{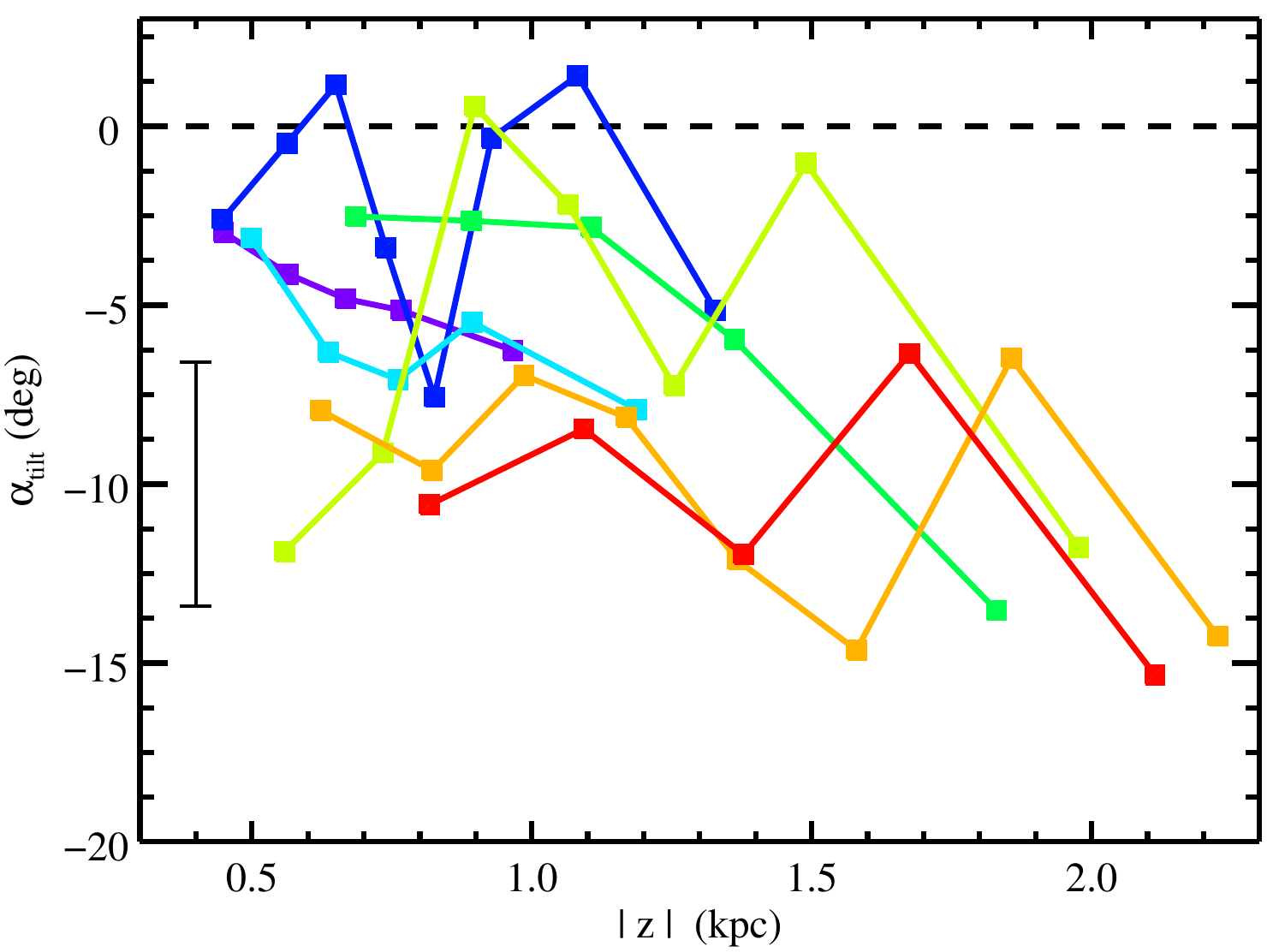}
    \includegraphics[width=0.45\linewidth]{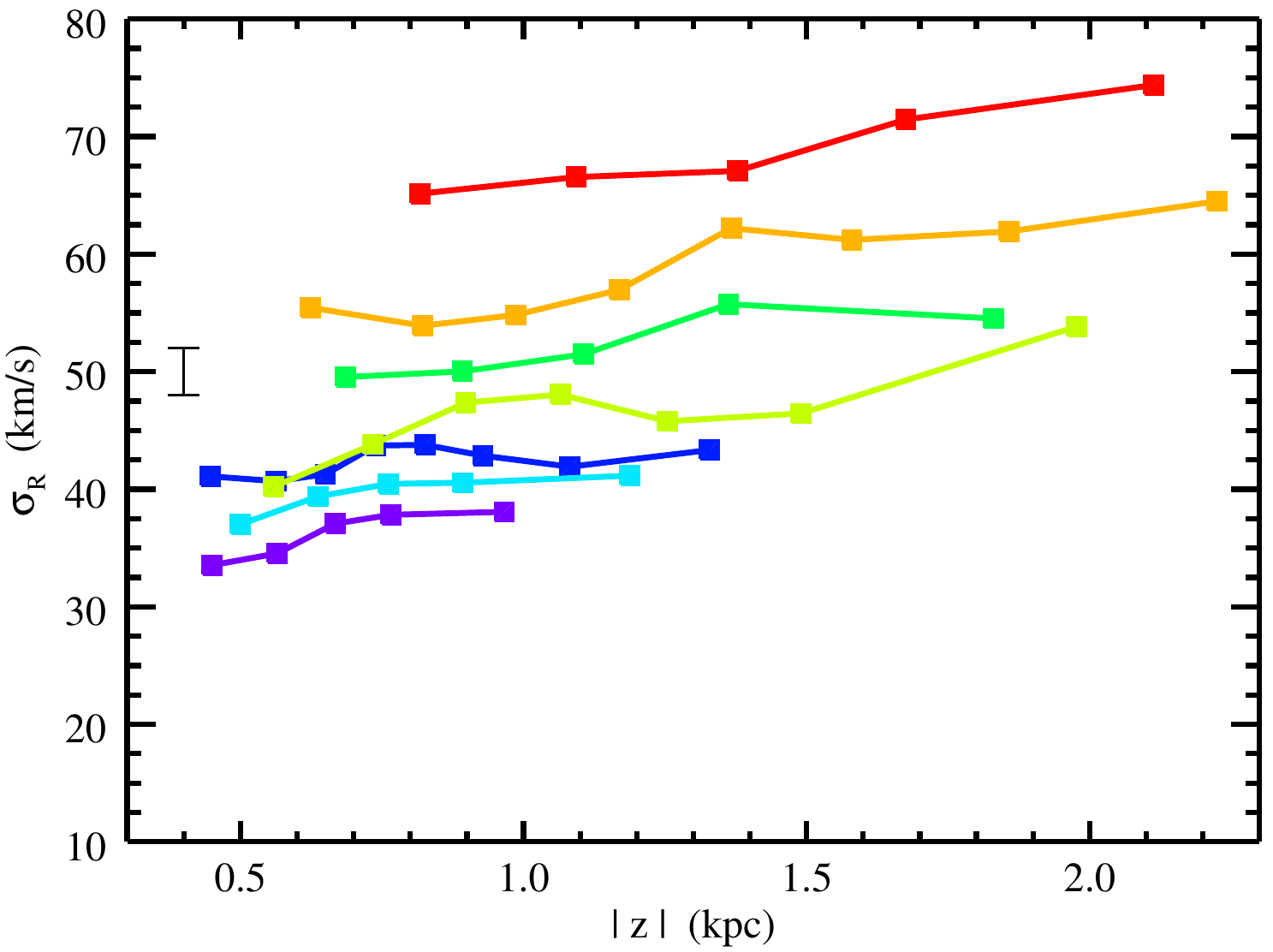}
    \includegraphics[width=0.45\linewidth]{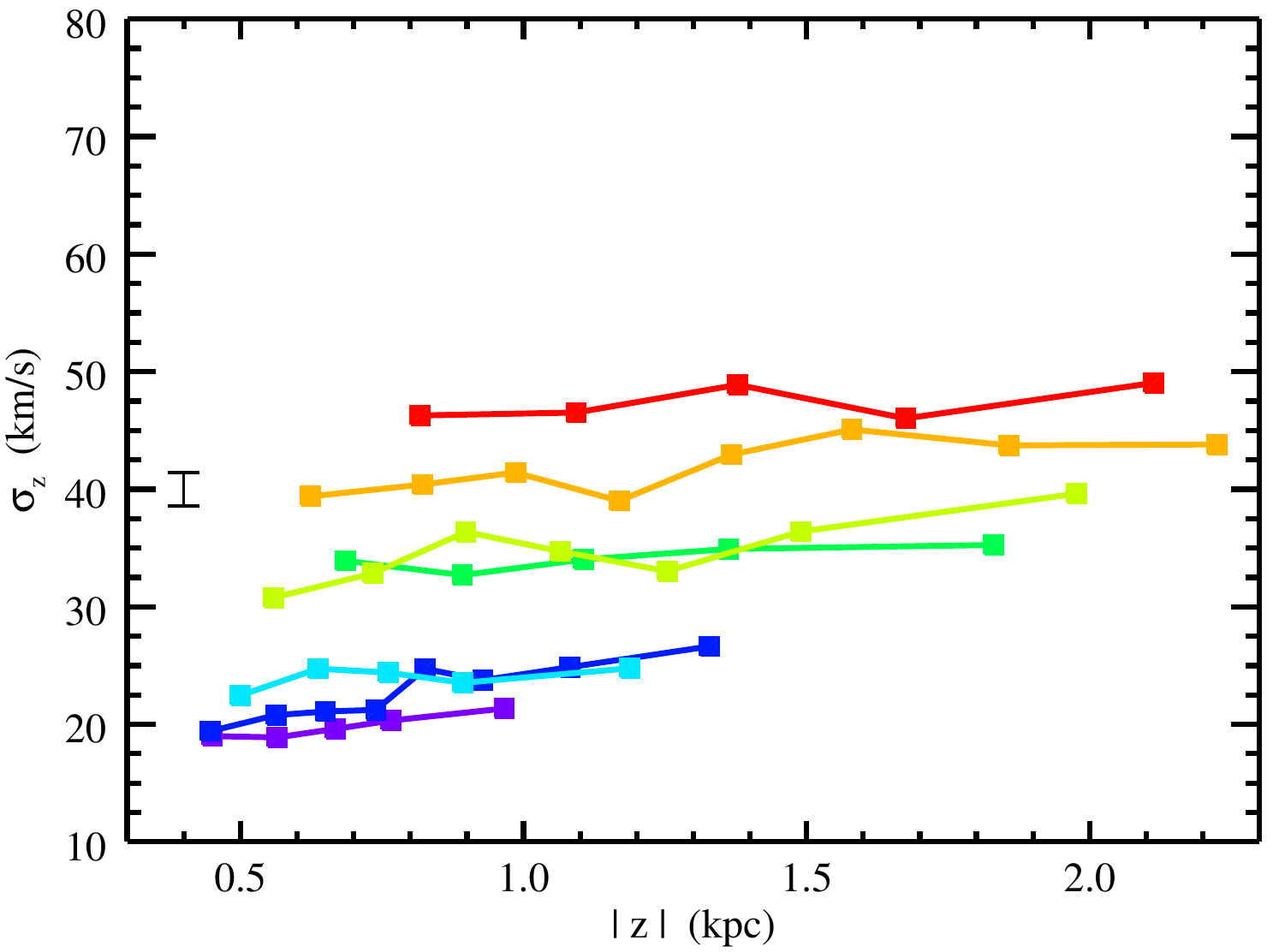}
    \caption{\emph{Top left:} The sub-division of SDSS/\SEGUE\ G-dwarf stars in the Solar neighbourhood according to their measured \afe\ abundance and \feh\ metallicity, with the number of stars per sub-sample indicated. Position in the \afe\--\feh\ plane can be used as a proxy for age; we reflect this in the colours, such that from purple to red the stars become older, on average.
\emph{Top right:} Non-zero tilt angle of the velocity ellipsoid for each sub-sample as function of height away from the Galactic mid-plane.
\emph{Bottom:} Nearly flat radial (left) and vertical (right) velocity dispersion as function of height for each sub-sample.
We provide the values of the above measurements in \autoref{tab:velellsubpop}.}
    \label{fig:velellsubpop}
\end{center}
\end{figure*}

As shown in the top-left panel of \autoref{fig:velellsubpop}, we divide our sample of G dwarfs into seven sub-samples in the plane of \afe{} versus \feh{}; we use a Voronoi binning scheme \citep{cappellari2003} to ensure comparable number of stars per sub-sample. We then sub-divide each sub-sample further in height $|z|$ away from the mid-plane so that each bin contains approximately 500 stars. This number of stars ensures that our MCMC discrete likelihood fits (see \autoref{SS:likelihoods}) yield robust results per bin on the three velocity ellipsoid components $\sigma_R$, $\sigma_z$ and $\overline{v_R v_z}$. In particular, an accurate measurement of the latter cross term is essential to infer the tilt angle $\atilt$ with a precision of $\lesssim 4$\dgr, indicated by the black error bar in the top-right panel of \autoref{fig:velellsubpop}. We provide the results from this analysis in \autoref{tab:velellsubpop}.

The corresponding uncertainties on the radial and vertical dispersions, shown in the bottom panels of \autoref{fig:velellsubpop}, are only $\lesssim 2$\kms. Although the dispersions change from bin to bin, within each sub-sample the dispersion is nearly constant with $|z|$, consistent with earlier findings of vertically near-isothermal behaviour of mono-abundance populations \citep[e.g.][]{liu2012,bovy2012c}. For the $\alpha$-older and more metal-poor stars with somewhat larger Voronoi bins, the remaining variation might be ascribed to a change with height in the relative contribution of stars with different kinematics. However, for the $\alpha$-younger and more metal-rich stars that are probing lower heights, a decrease in dispersion toward the mid plane is expected, but the amplitude will depend on the amount of dark matter (see also the solid and dashed curves in \autoref{fig:vert_results}) as well as the tilt of the velocity ellipsoid.

The top-right panel of \autoref{fig:velellsubpop} shows a clear non-zero tilt that increases in magnitude away from the mid-plane. Since the $\alpha$-older stars are typically probing larger heights, the assumption of decoupled radial and vertical motion in the above vertical Jeans analysis is likely to be more incorrect than for the $\alpha$-younger stars. So the inference that we made in \autoref{SS:vert_dispersions} -- that a gravitational potential with a significant presence of dark matter is more plausible -- is perhaps too premature; though we note that the velocity ellipsoid tilt is also significantly non-zero for the $\alpha$-younger stars, which casts doubt on our conclusions for that sub-sample as well. We have shown here that, within the measurement uncertainties, the tilt angle at a given height is consistent between the different sub-samples. Thus, henceforth, we shall consider the sample of G dwarfs together to improve the statistical precision on the measured velocity ellipsoid tilt.

\subsection{Tilt angle}
\label{SS:tiltangle}

\begin{figure*} \centering
    \includegraphics[width=1.0\textwidth]{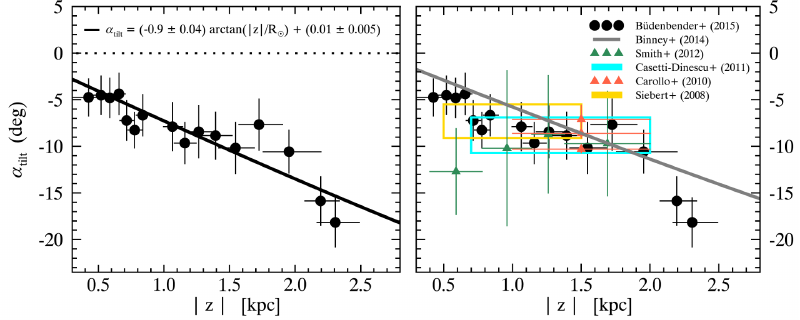}
    \caption{Tilt angle $\atilt$ of the velocity ellipsoid as function of height $|z|$ away from the mid-plane at the Solar radius. The filled circles are measurements with uncertainties indicated by the vertical error bars based on $\sim$1000 G-dwarf stars per bin in height with the bin-size indicated by the horizontal error bars.
    \emph{Left:} The tilt angle is significantly non-zero everywhere with best-fit arctan relation as indicated by the solid curve that is close to spherical alignment.
    \textit{Right:} Our tilt angle measurements are consistent with previous determinations, but significantly improved. We provide the measurements of the tilt angle as well as halo contamination fraction in \autoref{tab:tiltangle}.
    }
    \label{fig:tiltangle}
\end{figure*}

The left panel of \autoref{fig:tiltangle} shows the tilt angle $\atilt$ of the velocity ellipsoid as function of height $|z|$ away from the mid-plane at the solar radius. The measurements are based on our MCMC discrete likelihood fitting (see \autoref{SS:likelihoods}), with around 1500 G-type dwarf stars per bin in height. The vertical error bars indicate the standard deviation around the mean in the $\atilt$ values of the MCMC chain after convergence; the horizontal error bars indicate the size of the bin in $|z|$ around the median value (see also \autoref{tab:tiltangle}).

Over the full range in height probed from about 0.4 to 2.0\,kpc, the tilt angle is significantly non-zero and, thus, everywhere inconsistent with the assumption of decoupled radial and vertical motion. Whereas the latter would imply cylindrical alignment of the velocity ellipsoid, the measurements are instead consistent with a velocity ellipsoid pointing toward the Galactic centre: the solid curve represents the best-fit of the relation
\begin{equation}
  \atilt = (-0.90 \pm 0.04) \arctan(|z|/R_{\sun}) - (0.01 \pm 0.005),
  \label{eq:tiltanglefit}
\end{equation}
which is close to the case of alignment with the spherical coordinate system for which $\atilt = \arctan(|z|/R_{\sun})$.

In the case that the (local) potential is of separable St\"ackel form and axisymmetric, the velocity ellipsoid is aligned with the prolate spheroidal coordinate system \citep[e.g.][]{dezeeuw1985}. Expressed in cylindrical coordinates, the tilt angle is then given by
\begin{equation}
  \tan(2\atilt) = \frac{2 R z}{R^2 - z^2 + \Delta^2},
   \label{eq:tiltanglestaeckel}
\end{equation}
where $\Delta \ge 0$ is the focus of the prolate spheroidal coordinate system. The uncertainties in the tilt angle measurements allow for $\Delta/R_{\sun} \lesssim 0.24 (0.42)$ within $1\sigma$ ($3\sigma$) confidence limits, which includes the limiting case of spherical alignment with $\Delta=0$.

\subsection{Literature comparison}
\label{SS:literature}

In the right-panel of \autoref{fig:tiltangle}, we compare our estimate of the tilt angle as a function of distance from the mid-plane with estimates from previous studies.

\citet{siebert2008} used 580 red-clump stars below the Galactic mid-plane from the second data release of the RAdial Velocity Experiment (RAVE), to infer a tilt angle of $7.3 \pm 1.8$\dgr\ for heights $0.5 < |z|/\mathrm{kpc} < 1.5$. \citet{casetti-dinescu2011} combined data from the fourth release of the Southern Proper Motion Program and the same second release of RAVE for 1450 red-clump stars above and below the Galactic mid-plane to find a tilt angle of $8.6 \pm 1.8$\dgr\ for heights $0.7 < |z|/\mathrm{kpc} < 2.0$. After accounting for the flip in sign of $\atilt$ from below to above the Galactic mid-plane, \autoref{fig:tiltangle} shows that both measurements are consistent with our findings especially when taking into account the large range in heights around the mean $|z|\sim1$\,kpc.

Over a similar range in heights $1 < |z|/\mathrm{kpc} < 2$, \citet{carollo2010} found, based on a sample of more than ten thousand calibration stars from SDSS DR7, a consistent tilt angle of $7.1 \pm 1.5$\dgr\ for stars with metallicity $-0.8<$ \feh\ $<-0.6$, but a larger tilt angle of $10.3 \pm 0.4$\dgr\ for more metal-poor stars with $-1.5<$ \feh\ $<-0.8$. However, given that more metal-poor stars are relatively more abundant at larger heights, it is likely that both values are fully consistent with the $>10$\dgr\ change in tilt angle we find over this large range in height.
\citet{smith2012} also used SDSS DR7 data, but restricted to Stripe 82, to exploit the high-precision photometry and proper motions. They measured the tilt angle in four bins in the height range $0.5 < |z|/\mathrm{kpc} < 1.7$ for stars with metallicity \feh\ $<-0.6$ and more metal-poor stars with $-0.8<$ \feh\ $<-0.5$, and concluded that, despite larger uncertainties, the tilt angles are consistent with spherical alignment of the velocity ellipsoid; the few measurements that appear at larger (negative) tilt angles they believe to be an artefact.

Recently, \citet{binney2014} used $>400,000$ stars from the fourth data release of RAVE to infer, under the assumed tilt angle variation $\atilt \propto \arctan(|z|/R_{\sun})$, a proportionality constant of $\sim0.8$ except for hot dwarfs with $\sim0.2$. The former gradient is consistent with our measurements in \autoref{fig:tiltangle} and the corresponding best-fit relation given in equation~(\ref{eq:tiltanglefit}), but the hot-dwarfs gradient appears too shallow, although a more quantitative comparison is unfortunately not possible due to missing uncertainties on the inferred gradients.

\begin{table}
    \caption{Measured tilt angle (in degrees) as function of height in pc from \autoref{fig:tiltangle}. The last column shows halo contamination fraction (in \%). Their errors are estimated from the standard-deviations of the post-burn parameter distributions. }
    \label{tab:tiltangle}
    \vertspace{1.1}
    \begin{tabular}{@{}ccc|ccc@{}}
        \toprule
        z & $\atilt$ & $\epsilon$ &
        z & $\atilt$ & $\epsilon$ \\
        (pc) & (deg) & (\%) &
        (pc) & (deg) & (\%) \\
        \midrule
 425 &  -4.70 $\pm$ 2.00 &   2.8 $\pm$ 0.8  &    1159 &  -9.60 $\pm$ 2.20 &   3.6 $\pm$ 1.1 \\
 520 &  -4.50 $\pm$ 2.10 &   3.3 $\pm$ 0.9  &    1265 &  -8.40 $\pm$ 2.80 &   4.7 $\pm$ 1.2 \\
 587 &  -4.80 $\pm$ 2.10 &   4.1 $\pm$ 1.0  &    1393 &  -8.80 $\pm$ 2.50 &   4.7 $\pm$ 1.4 \\
 657 &  -4.40 $\pm$ 2.20 &   3.4 $\pm$ 0.9  &    1545 & -10.20 $\pm$ 2.80 &   3.7 $\pm$ 1.4 \\
 715 &  -7.20 $\pm$ 2.10 &   2.5 $\pm$ 0.8  &    1724 &  -7.70 $\pm$ 2.70 &   4.2 $\pm$ 1.3 \\
 777 &  -8.30 $\pm$ 1.90 &   2.4 $\pm$ 0.8  &    1953 & -10.60 $\pm$ 2.30 &   5.7 $\pm$ 1.5 \\
 838 &  -6.70 $\pm$ 2.20 &   3.2 $\pm$ 0.9  &    2194 & -15.90 $\pm$ 2.60 &   9.4 $\pm$ 3.1 \\
1064 &  -7.90 $\pm$ 2.60 &   3.0 $\pm$ 1.1  &    2306 & -18.20 $\pm$ 2.60 &   6.7 $\pm$ 2.9 \\
      \bottomrule
    \end{tabular}
\end{table}

\section{Discussion and conclusion}
\label{S:discconcl}

In this paper, we have accurately measured the velocity ellipsoid of the Milky Way disk near the Sun. To do this, we used a well-characterised sample of $>$16,000 G-type dwarf stars from the \SEGUE\ survey and fit their discrete kinematic data using a likelihood method that accounts for halo star contaminants. In combination with Markov Chain Monte Carlo (MCMC) sampling, we have robustly measured the velocity ellipsoid components as a function of height away from the Galactic mid-plane for a number of chemically-distinct sub-samples.

To begin, we separated our sample into two sub-samples based on their metallicity and $\alpha$-element abundances. As these sub-samples are tracers of the same underlying gravitational potential, fitting Jeans models to the vertical density and dispersion profiles for each sub-sample independently should yield the same constraint on the local dark matter density. Instead, we found large variations: metal-rich, low-$\alpha$-abundant stars require a significant local dark matter density, while metal-poor, high-$\alpha$-abundant stars do not need any dark matter. As the latter stars are relatively more abundant at larger vertical heights, we believe this is the consequence of a coupling between vertical and radial motions that becomes stronger with height. In turn, this should be detectable as an increase in the tilt angle of the velocity ellipsoid with height.

Next, we measured the velocity ellipsoid components in the meridional plane as function of height, for seven chemically-distinct sub-samples. We found radial and vertical dispersions, $\sigma_R$ and $\sigma_z$, that are approximately constant with height, consistent with the isothermally profiles found in earlier studies \citep[e.g.][]{liu2012, bovy2012c}. Between the sub-samples, the amplitudes of both $\sigma_R$ and $\sigma_z$ increase when the stars are less metal-rich and more $\alpha$-abundant, in line with the age-velocity relation observed in the Solar neighbourhood \citep[e.g.][]{casagrande2011}. The cross term $\mvRz$ together with $\sigma_R$ and $\sigma_z$ yields a tilt angle of the velocity ellipsoid that is clearly non-zero and its amplitude indeed increasing with height.

\begin{figure*}
 \begin{center}
    \includegraphics[width=0.45\linewidth]{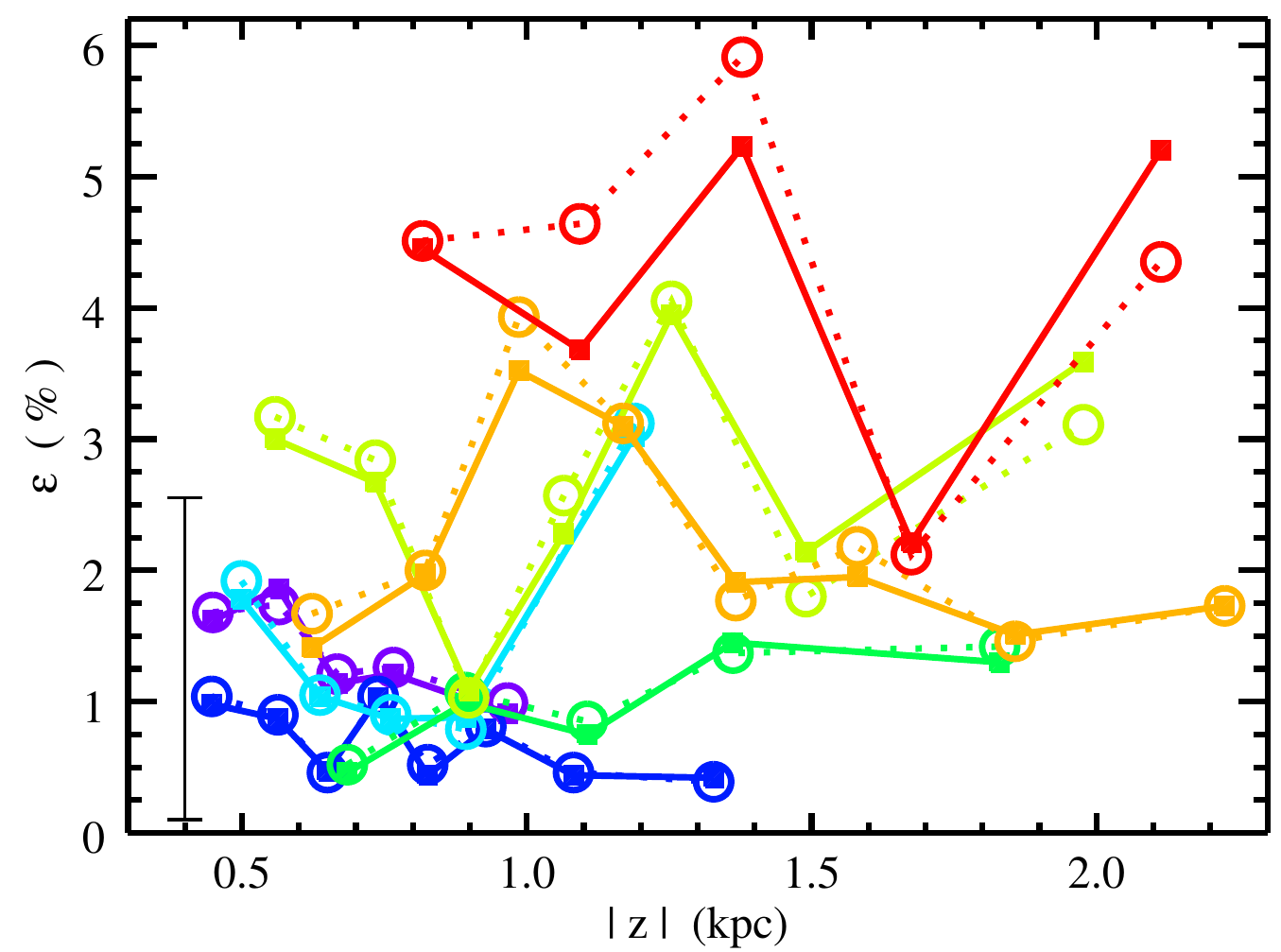}
    \includegraphics[width=0.45\linewidth]{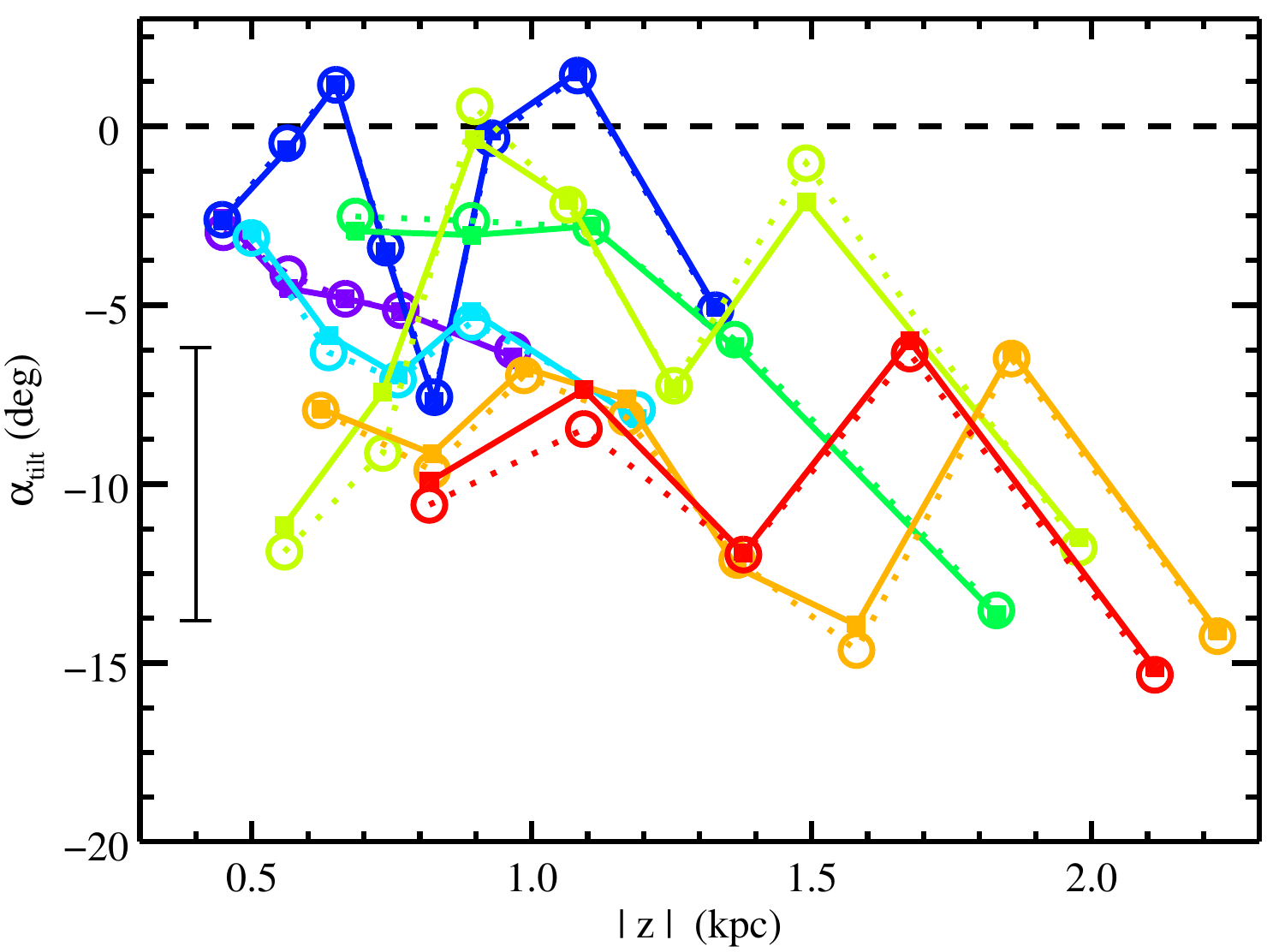}
 \end{center}
\caption{Halo contamination fraction (left panel) and tilt angle (right panel) as a function of distance from the Galactic plane for two different contamination models. The colours reflect the sub-samples as illustrated in the top-left panel of \autoref{fig:velellsubpop}. The square symbols joined with solid lines use the inner-component of the dual-halo model described in \citet{carollo2010} to describe the expected halo contamination; for comparison, we show the canonical single-component halo model from \citet{schoenrich2011} that we adopted for this study as the open circles joined with dashed lines. We see that the contamination fractions are very similar, regardless of the particular contamination model used, and, in turn, that the effect on the tilt angles inferred is minimal. In both panels, any differences are well within the uncertainties indicated by the error bars.}
\label{fig:compinnerhalo}
\end{figure*}

For modelling the contamination by halo stars, we adopted a canonical single-component Galactic halo \citep[e.g.][]{schoenrich2011}. However, several studies have suggested a two-component halo structure \citep[e.g.][]{carollo2007, carollo2010, deJong2010, nissen2010, kinman2012, an2013, hattori2013}. \citet{carollo2007, carollo2010} showed that the outer-halo component is only dominant beyond $\sim$15-20 kpc and at metallicities \feh $< -$2.0; as our sample does not extend above 3 kpc in height, we expect negligible contamination from this component. Nevertheless, to ensure that our results are not sensitive to the particular choice of contamination model, we repeat our calculations using the inner component of the dual-halo model from \citet{carollo2010}; that is, we use dispersions $\sigma_\mathrm{R,halo} = 150 \pm 2$~\kms\ and $\sigma_\mathrm{z,halo} = 85 \pm 1$~\kms\ in equation~\eqref{eq:lih}. We show the results of this test in \autoref{fig:compinnerhalo}; the left panel shows the contamination fraction and the right panel shows the tilt angle, both as a function of distance from the Galactic plane. In both cases, the differences in the results from the two different halo models are well within the uncertainties and, thus, the results from our vertical Jeans models remain unchanged.

\begin{figure*}
 \begin{center}
    \includegraphics[width=0.45\linewidth]{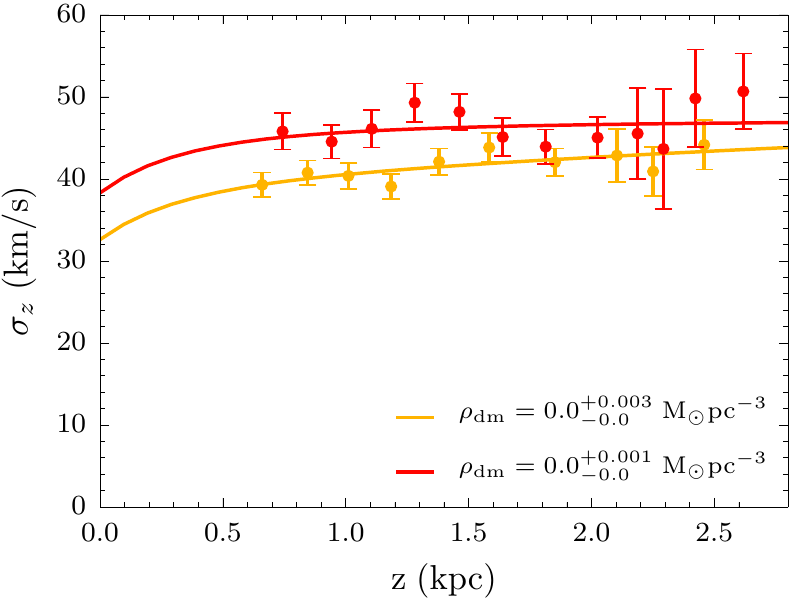}
    \includegraphics[width=0.45\linewidth]{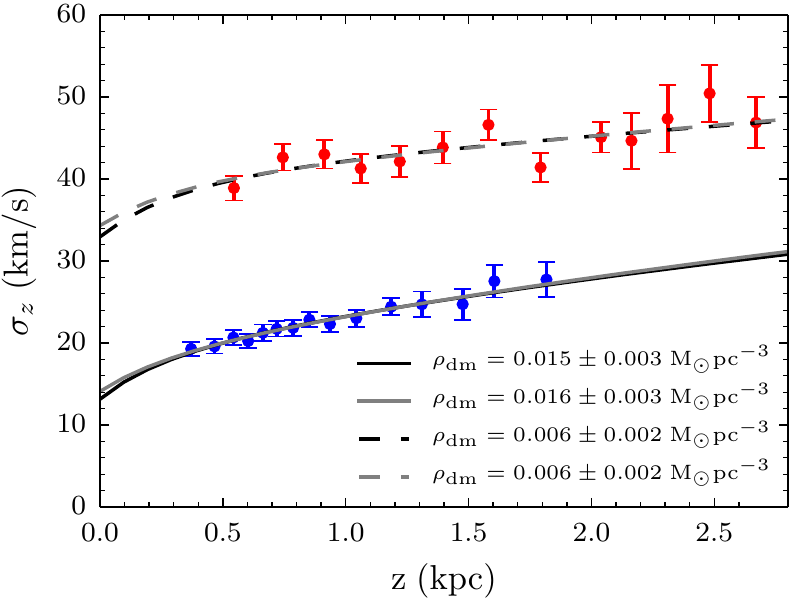}
 \end{center}
 \caption{Vertical velocity dispersion as function of height; similar to \autoref{fig:vert_results} where $\alpha$-young and $\alpha$-old stars are separately fitted by vertical Jeans models. The plots explore the effect of the metal-weak thick disk on our vertical Jeans models.
\textit{Left}: The orange and red dispersions are extracted from the subsamples coloured orange and red in \autoref{fig:velellsubpop}. The red subsample is more metal-poor and so more susceptible to the presence of metal-weak thick disk stars. The shapes of the two fitted profiles are very similar, as are the inferred local dark matter densities, indicating that the results obtained for the $\alpha$-old sample are largely insensitive to metallicity (see also \autoref{fig:velellsubpop}) and, hence, to the presence of a metal-weak tail to the thick disk.
\textit{Right}: Dispersion profiles for the same $\alpha$-young (blue) and $\alpha$-old (red) subsamples as shown in the bottom-right panel of \autoref{fig:vert_results}. Dashed lines show fits to the $\alpha$-old subsample and solid lines show fits to the $\alpha$-young subsample. The black lines are the original fits, also shown in \autoref{fig:vert_results}. The grey lines are fits with the metal-weak thick disk explicitly added to the gravitational potential. Once again, there is no significant difference in the fits or the inferred local dark matter density.}
 \label{fig:vertJ_mwtd}
\end{figure*} 

Some studies have also found indications of a metal-weak tail of the thick disk \citep[MWTD; e.g.][and references therein]{chiba2000}. If MWTD stars are present in our sample, they would be found among the $\alpha$-old thick-disk stars, primarily at lower metallicities. To test the effect they have on our results, we repeat the analysis from \autoref{SS:vert_dispersions} using the red and orange sub-samples selected in \autoref{fig:velellsubpop} as these two subsamples have similar (high) $\alpha$ abundances but different metallicities. We show the results of this test in the left panel of \autoref{fig:vertJ_mwtd}: the more metal-poor red subsample is shown as red points and the fitted Jeans model as a red line; the more metal-rich orange subsample is shown as orange points and the fitted Jeans model as an orange line. We see that the shapes of the two fitted profiles are very similar, indicating that the presence of any MWTD stars will have a negligible effect on the vertical Jeans model results.

Even so, we perform a further test to verify this expectation: we repeat the full \autoref{SS:vert_dispersions} analysis, but now consider the MWTD as an independent stellar population as suggested by \citet{carollo2010} and include a second thick disk component with exponential scale height 1.3~kpc and scale length of 2~kpc, in addition to a thick disk with 0.51\,kpc and 2.2\,kpc as exponential scale height and length. We show the results of this test in the right panel of \autoref{fig:vertJ_mwtd}. As before, the red and blue points show the dispersion profiles calculated from the $\alpha$-old and $\alpha$-young subsamples. The black lines show the original fits from \autoref{SS:vert_dispersions} and the grey lines show the new Jeans model fits with the MWTD explicitly included. It is clear that adding the extra MWTD component does not change our results. This further supports our conclusion that the differences we see between the $\alpha$-old and $\alpha$-young subsamples is due to the incorrect assumption regarding the separability of the radial and vertical motions and not because of missing components in our Galactic model.

As the tilt angle measurements between the sub-samples are fully consistent within the error bars, we were able to decrease the statistical uncertainties by combining all G dwarfs. This yields a tilt angle as function of height that is consistent with previous determinations, but significantly improved. The resulting measurements given in Table~\ref{tab:tiltangle} are very well fitted by the the relation  $\atilt = (-0.90 \pm 0.04) \arctan(|z|/R_{\sun}) - (0.01 \pm 0.005)$, which is close to alignment with the spherical coordinate system and hence a velocity ellipsoid pointing to the Galactic centre.

In the case of a St\"ackel potential, the tilt of the velocity ellipsoid is directly coupled to the shape of the gravitational potential and thus \emph{must} be the same for any sub-sample. In case of oblate axisymmetry the velocity ellipsoid is then aligned with the prolate spheroidal coordinate system. The resulting expression for the tilt angle (eq.~\ref{eq:tiltanglestaeckel}) can describe the tilt angle measurements as long as the focus of the latter coordinate systems is significantly smaller than the solar radius. Even if the St\"ackel potential is only a good approximation locally, this brings a convenient, and often fully analytical, expression of dynamical aspects that otherwise, even numerically, are very hard to achieve. One such example is the use of a local St\"ackel approximation to infer the integral of motions or actions \citep{binney2012}.


In a forthcoming paper, we obtain a solution of the axisymmetric Jeans equations along curvilinear coordinates that allows us to construct in a computationally efficient way models that allow for a non-zero tilt of the velocity ellipsoid. In this way, we can overcome the assumption of decoupled motion in the vertical Jeans models, while still being able to do a discrete likelihood fit with MCMC parameter inference, even for many thousands of stars at the same time. Among other benefits, this will enable a much more accurate measurement of the local dark matter density, especially with upcoming data from Gaia and spectroscopic follow-up surveys such as Gaia-ESO \citep{gilmore2012} and 4MOST \citep{deJong2012}.


\section*{Acknowledgements}
We thank Chao Liu for his support and feedback on the data analysis, and the referee for constructive comments. This work was supported by Sonderforschungsbereich SFB 881 ``The Milky Way System" (subproject A7) of the German Research Foundation (DFG).


\bibliographystyle{mn2e}
\bibliography{refs}


\appendix

\section{Effect of non-axisymmetry on tilt angle}
\label{A:nonaxi}

\begin{figure*}
\begin{center}
    \includegraphics[width=0.32\linewidth]{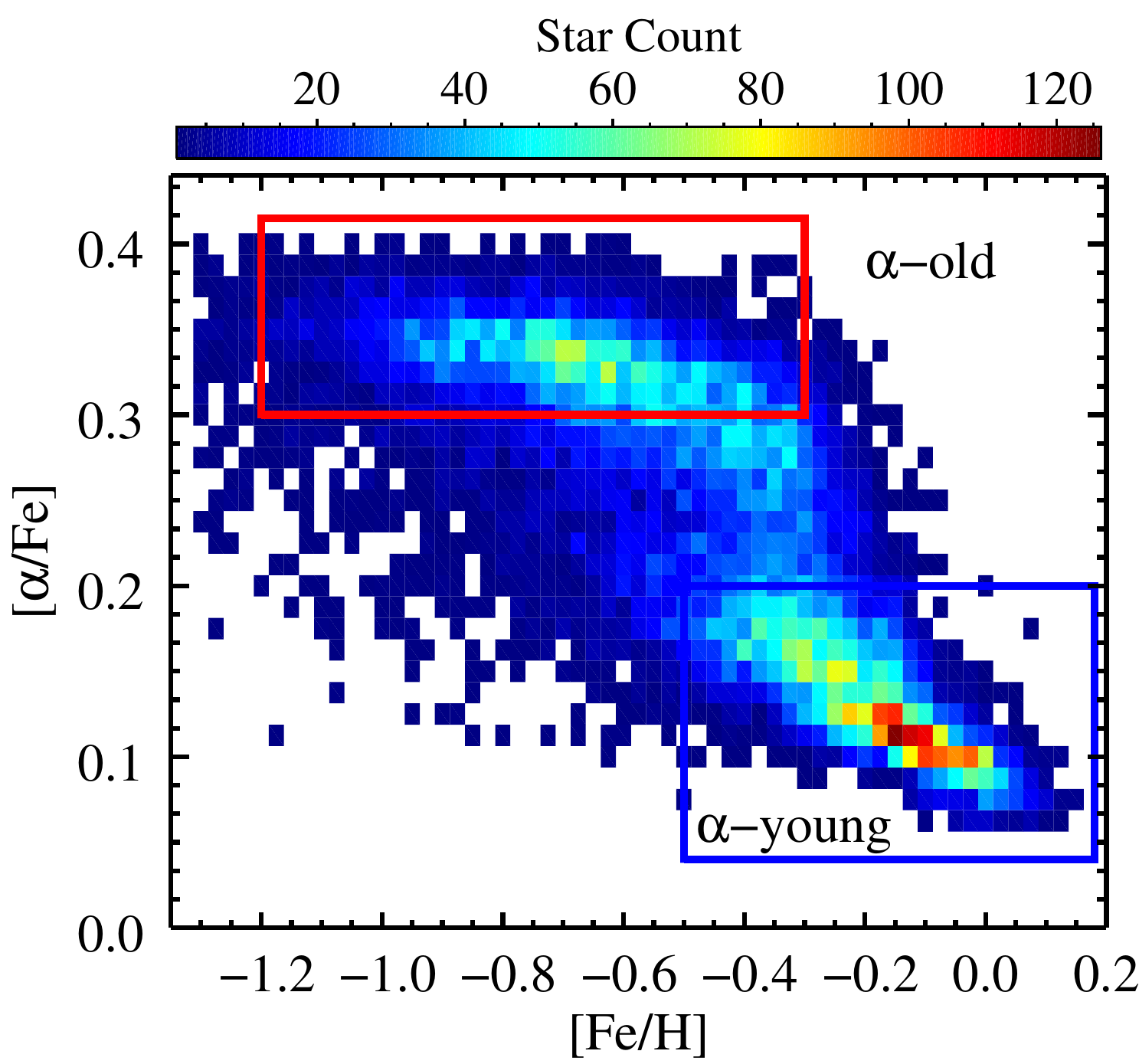}
    \includegraphics[width=0.32\linewidth]{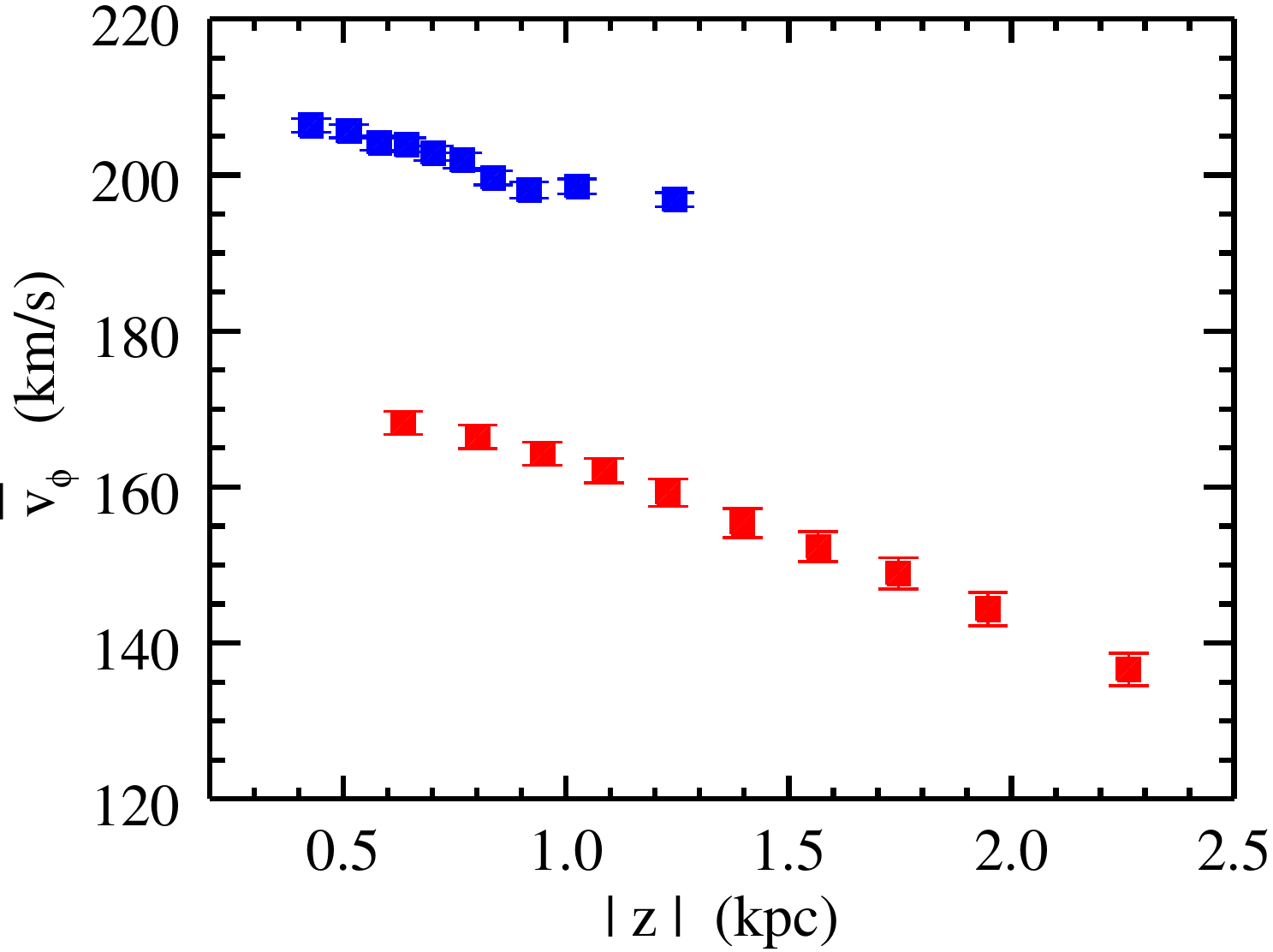}
    \includegraphics[width=0.32\linewidth]{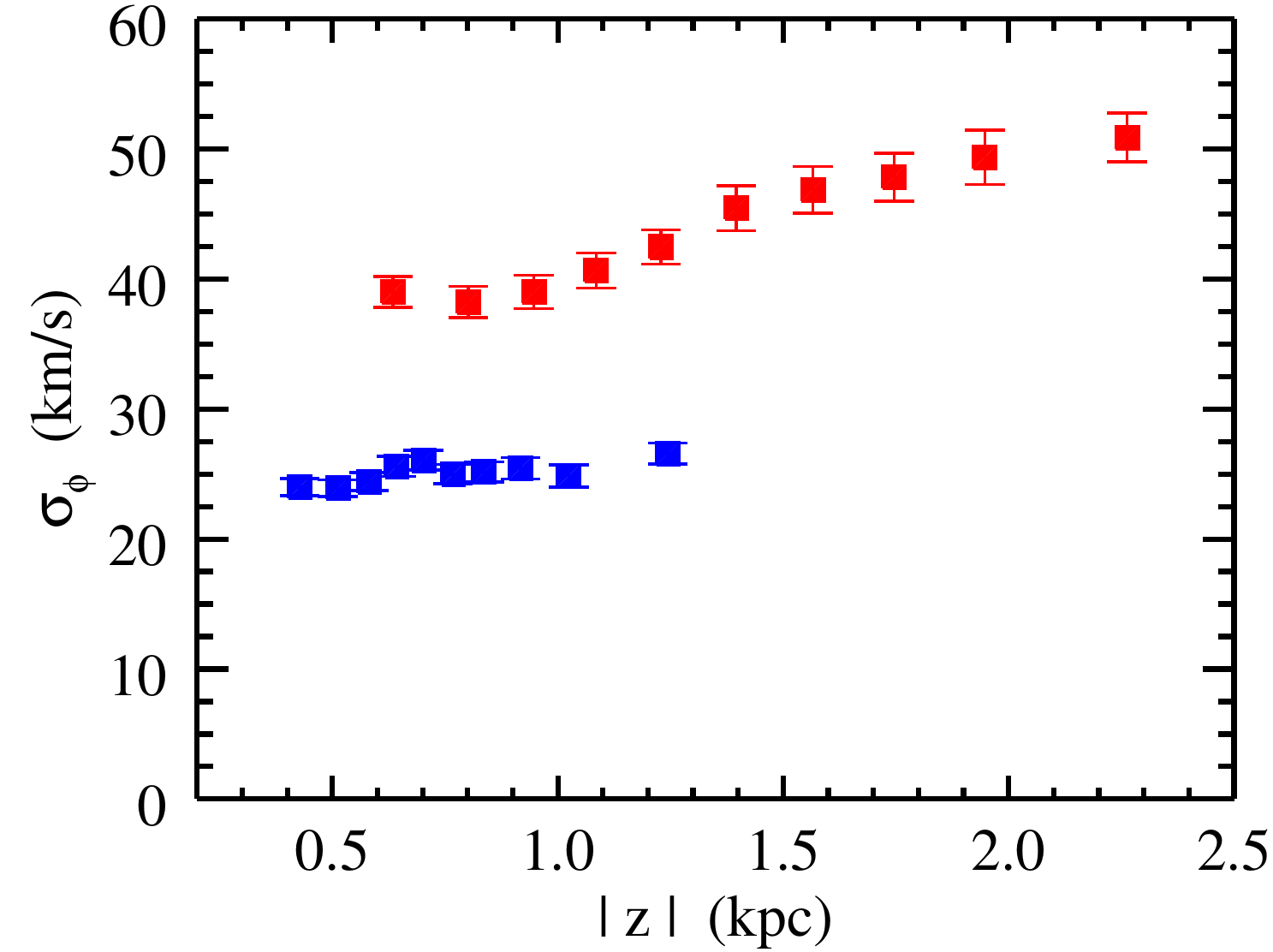}
    \includegraphics[width=0.32\linewidth]{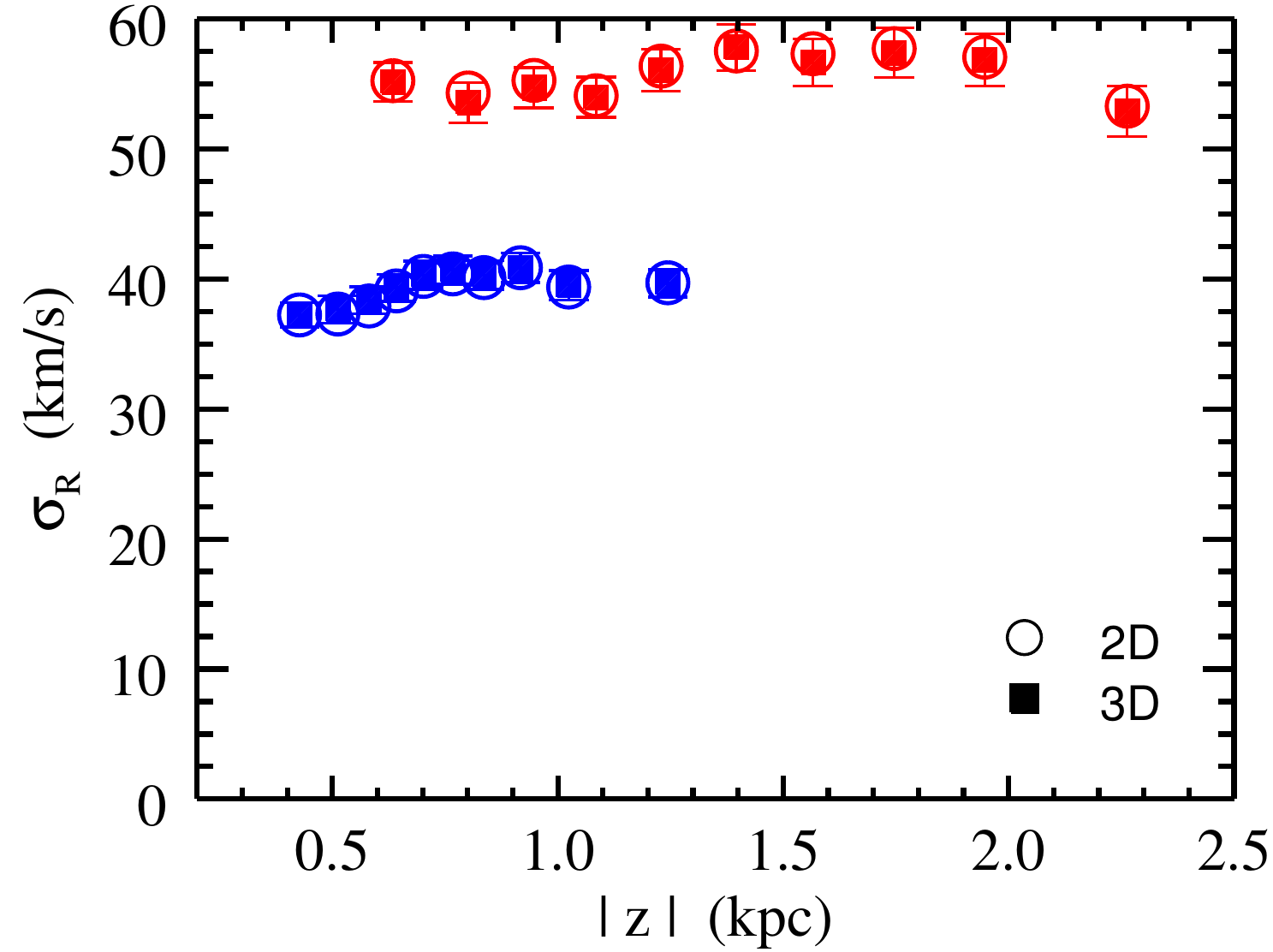}
    \includegraphics[width=0.32\linewidth]{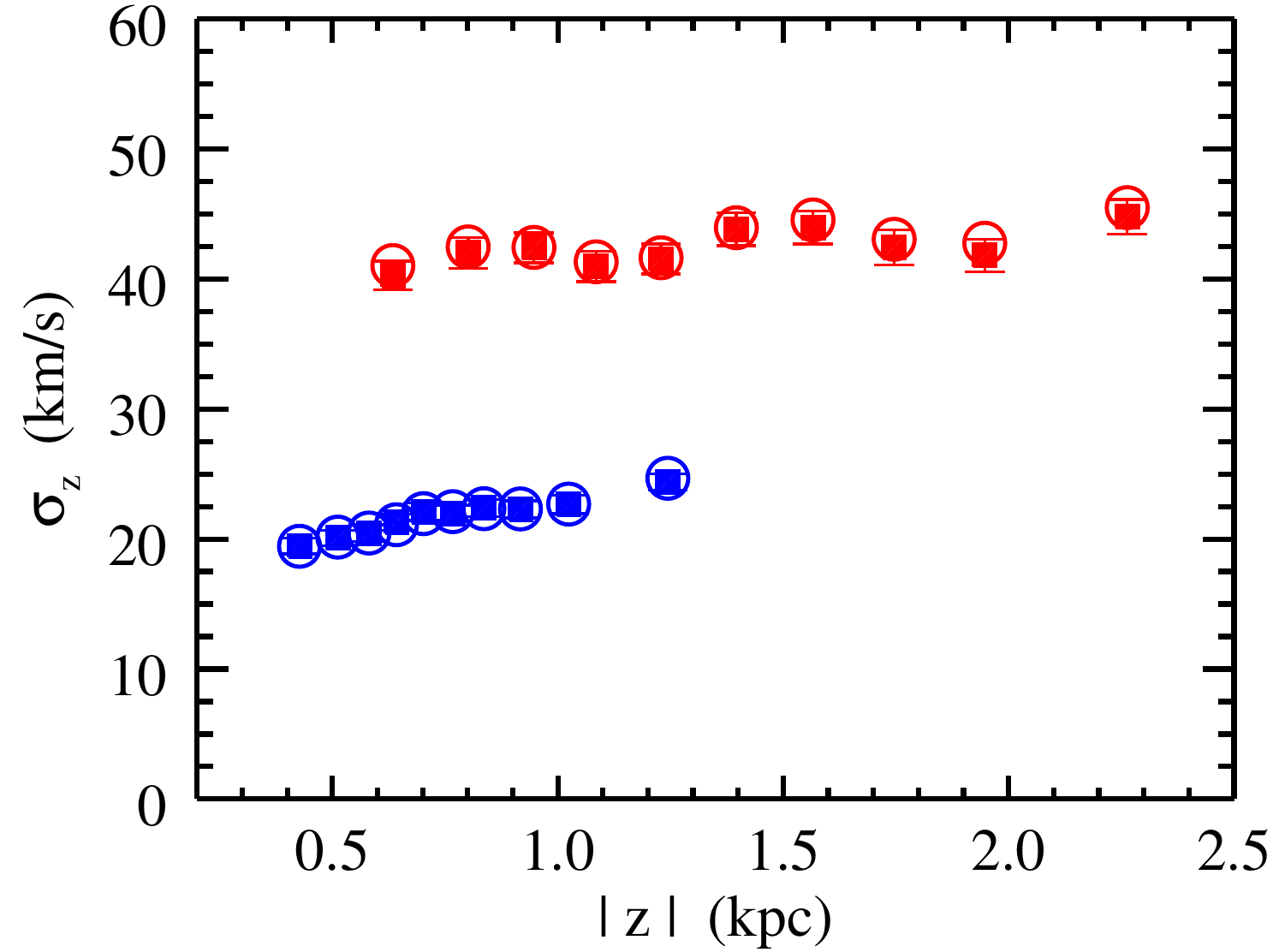}
    \includegraphics[width=0.32\linewidth]{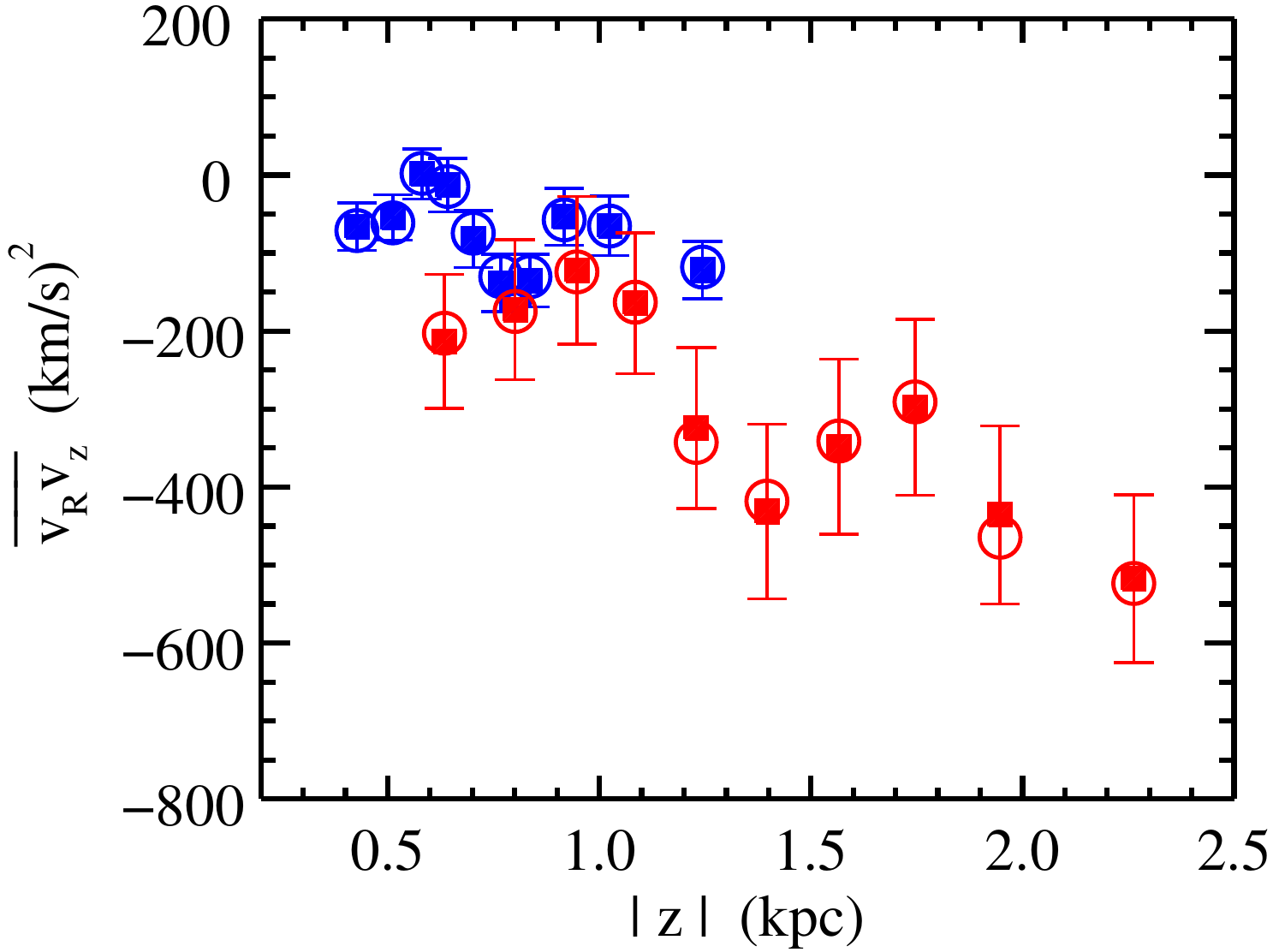}
    \caption{Top left: \afe\ abundances and \feh\ metallicities of the G-dwarf stars, identical to \autoref{fig:vert_results}. The red and blue boxes show the selections for the $\alpha$-old and $\alpha$-young sub-samples, respectively. These same colours are used in all other panels. Top middle and right: Azimuthal mean velocity and velocity dispersion as function of height $|z|$ away from the mid-plane at the Solar radius. Bottom row: Radial and vertical velocity dispersion and their correlated second velocity moment for the two sub-samples. The open symbols show the results for the multivariate Gaussian velocity distribution of rank 2, while the filled symbols show the corresponding results of a multivariate Gaussian of rank 3.}
    \label{fig:2dvs3d}
\end{center}
\end{figure*}

As described in \autoref{SS:meridionalplane}, the tilt of the velocity ellipsoid is independent of the azimuthal velocity in case of axisymmetry. In the bottom panels of \autoref{fig:2dvs3d}, we show that excluding or including $v_\phi$ yields consistent results for the velocity ellipsoid components in the meridional plane, $\sigma_R$, $\sigma_z$ and $\mvRz$, that make up the title angle.
For an $\alpha$-old (red) and an $\alpha$-young (blue) sub-sample selected as indicated in the top-left panel, the open circles adopt a multivariate Gaussian of rank 2 in the likelihood fitting described in \autoref{SS:likelihoods}, while the filled squares include the azimuthal velocities in the fit by adopting a multivariate Gaussian of rank 3. The inferred values are nearly indistinguishable, so that including $v_\phi$ is not needed and actually and would lead to slightly larger uncertainties as well as the complication that the distribution in $v_\phi$ is typically non-Gaussian.
Even so, the inferred azimuthal mean velocity $\mvp$ and velocity dispersion $\sigma_\phi$, shown in the top-middle and top-right panel, are as expected for a dynamical warmer $\alpha$-old sub-sample with $\mvp/\sigma_\phi$ smaller than an dynamically colder $\alpha$-younger sub-sample.

\begin{figure*}
\begin{center}
    \includegraphics[width=0.32\linewidth]{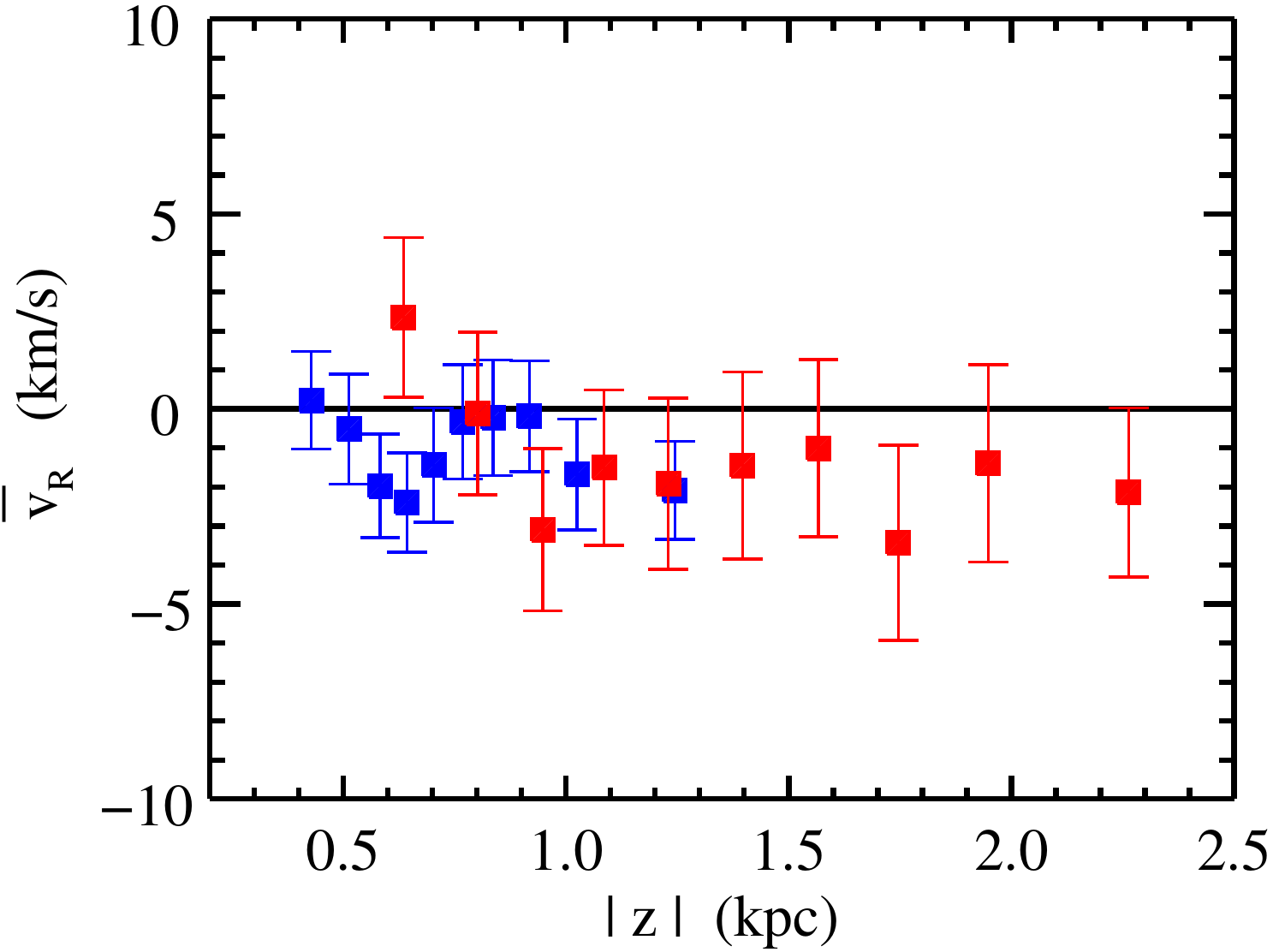}
    \includegraphics[width=0.32\linewidth]{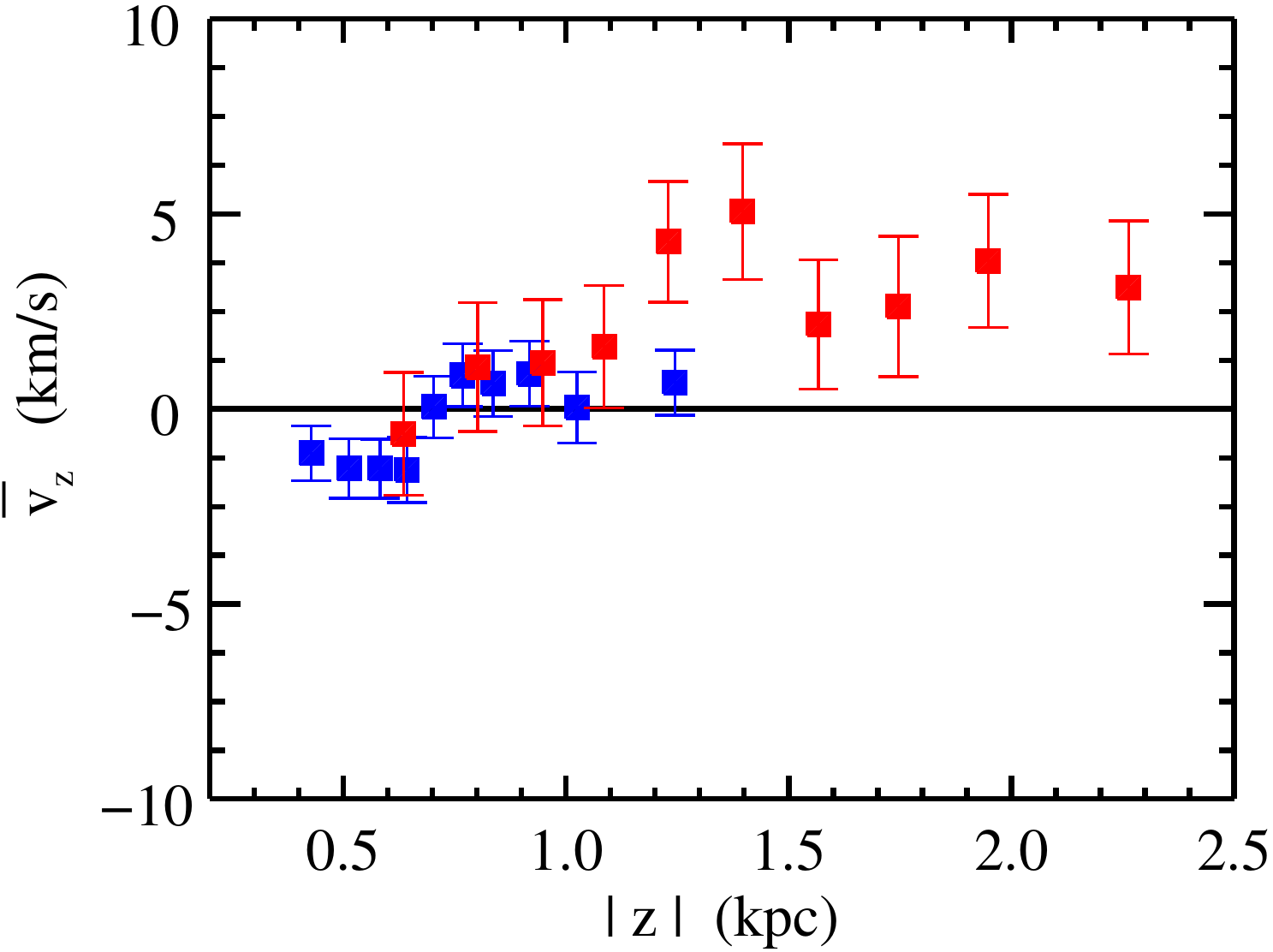}
    \includegraphics[width=0.32\linewidth]{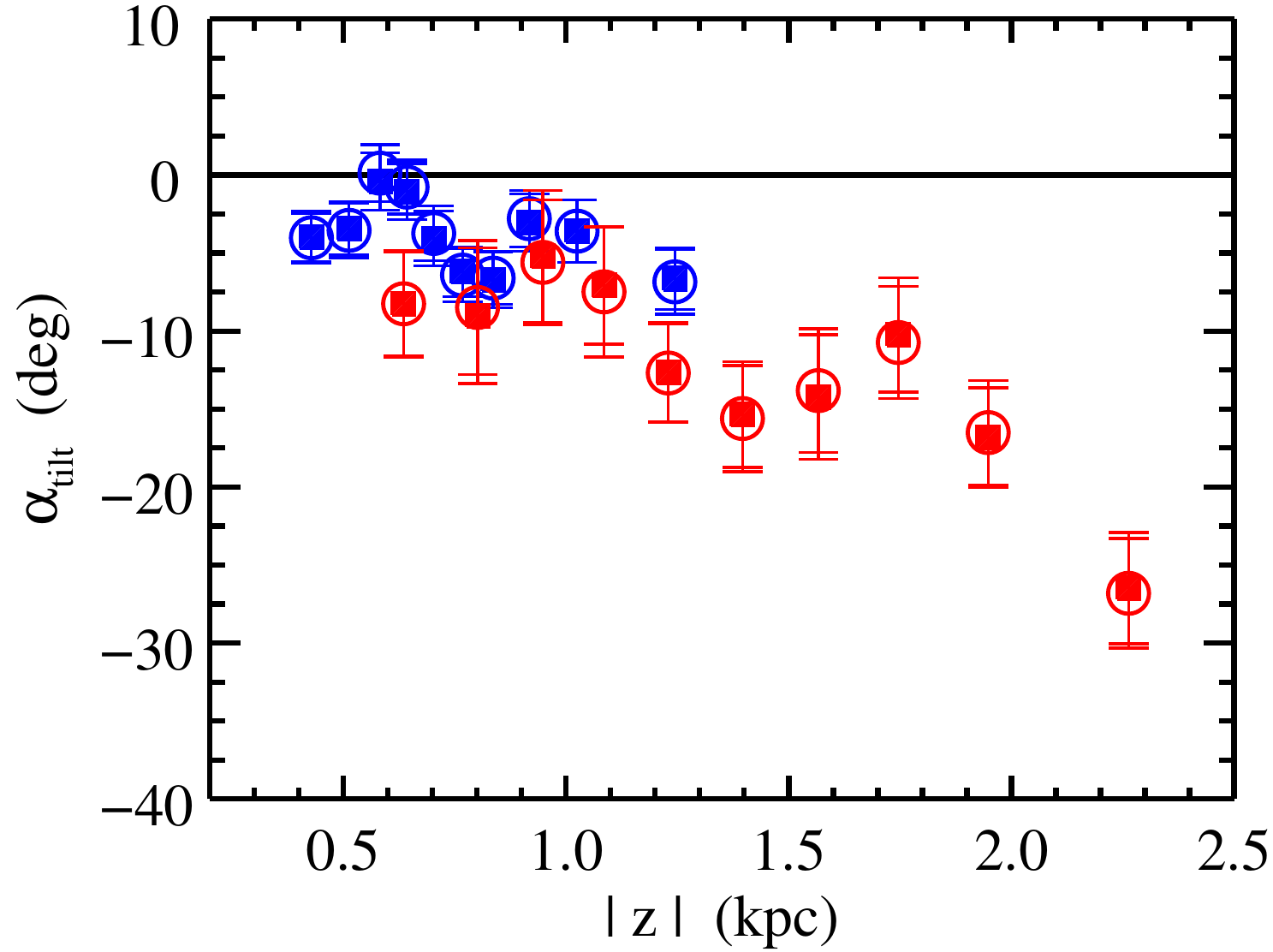}
    \includegraphics[width=0.32\linewidth]{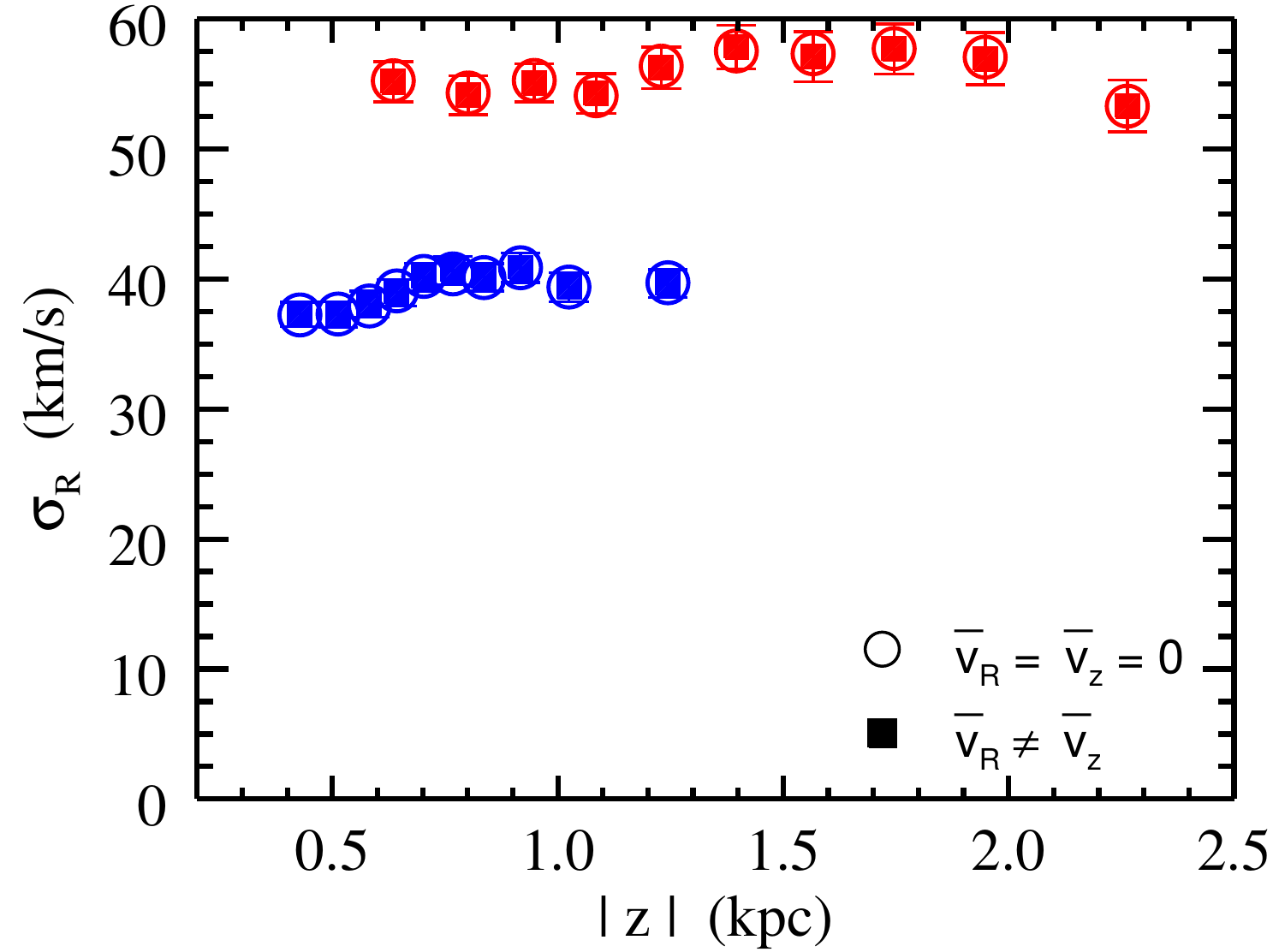}
    \includegraphics[width=0.32\linewidth]{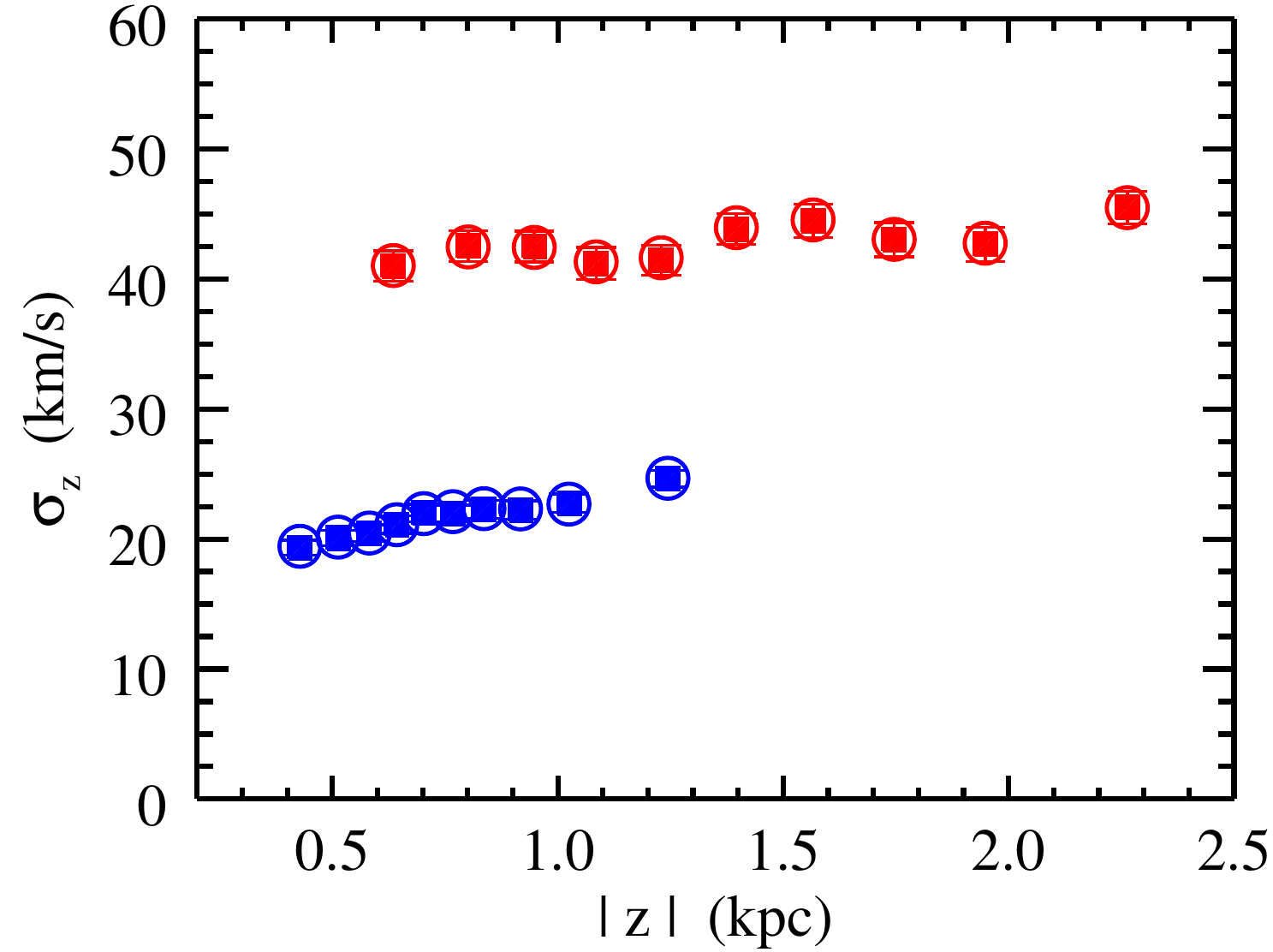}
    \includegraphics[width=0.32\linewidth]{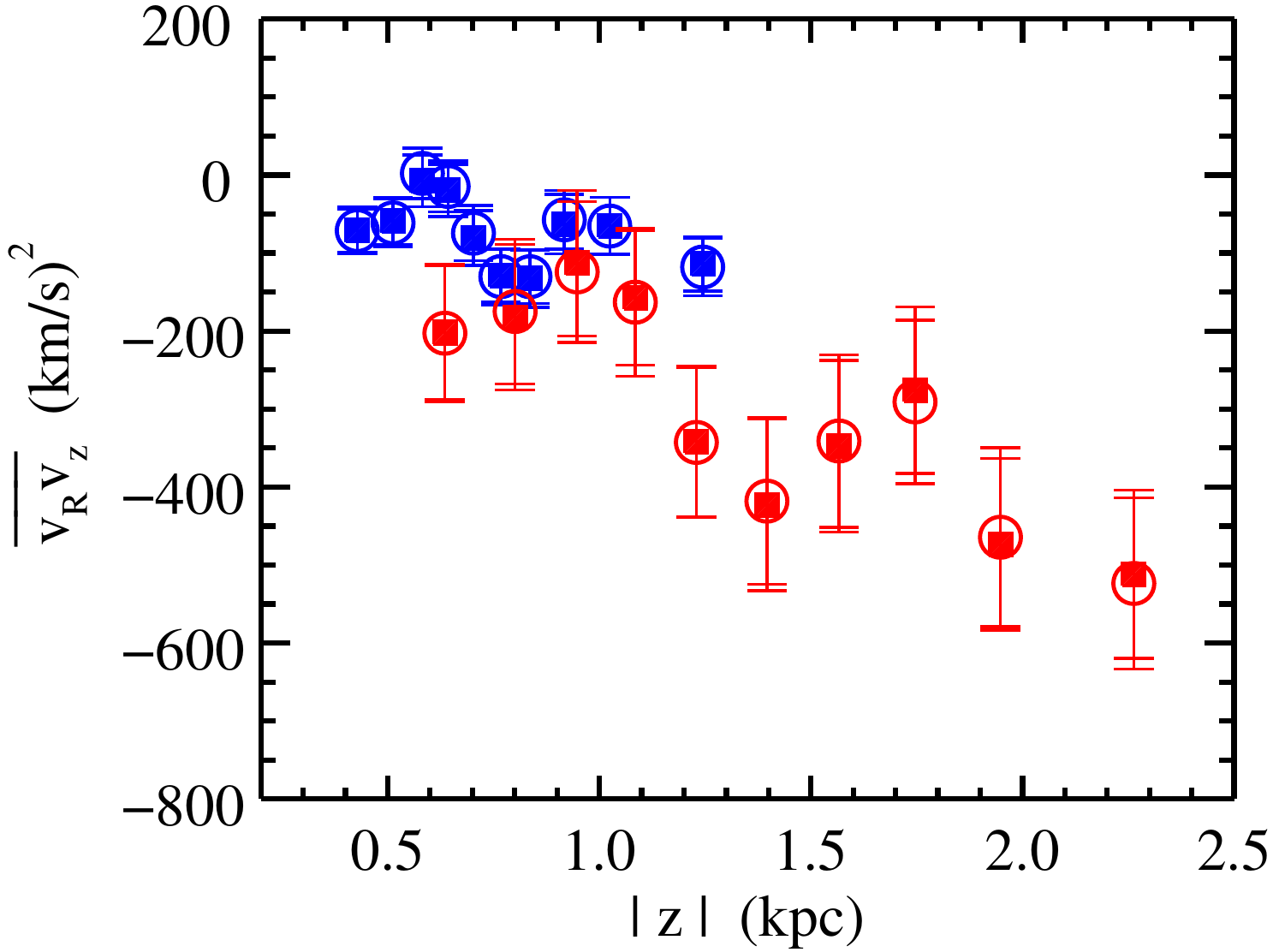}
    \caption{Dynamical profiles for the $\alpha$-old (red) and $\alpha$-young (blue) as a function of distance from the mid-plane at the solar radius. Top left: mean radial velocity. Top middle: mean vertical velocity. Top right: tilt angle of the velocity ellipsoid. Bottom left: radial velocity dispersion. Bottom middle: vertical velocity dispersion. Bottom right: correlated second velocity moment. In the latter four panels, the open symbols show the case for which we assume $\overline{v_R} = \overline{v_z} = 0$ and the filled symbols show the case where $\overline{v_R}$ and $\overline{v_z}$ are free parameters in the likelihood function (equation~\ref{eq:lijd}).}
    \label{fig:nonzeromeanvel}
\end{center}
\end{figure*}

Restricting to the meridional plane, the mean radial and vertical motion are zero in case of axisymmetry and hence should not effect the tilt angle. In \autoref{fig:nonzeromeanvel}, we show that even though $\overline{v_R}$  and $\overline{v_z}$ are observed to be mildly non-zero there is no significant effect on the velocity ellipsoid components and corresponding tilt angle.
For the same $\alpha$-old (red) and an $\alpha$-young (blue) sub-sample as in \autoref{fig:2dvs3d}, the open circles show the latter quantities measured in case we set $\overline{v_R} = \overline{v_z} = 0$, while in case of the filled squares the means of the bivariate Gaussians are free parameters. The measured velocity ellipsoid components and corresponding tilt angle are again nearly indistinguishable, so that the means of the bivariate Gaussians can be safely set to zero; the number of free parameters are reduced, so that the statistical uncertainty on particular $\mvRz$ and thus also the tilt angle decrease.
When left free, both $\overline{v_R}$  and $\overline{v_z}$ show small but significant deviations of a few \kms\ from zero, consistent with earlier findings \citep[e.g.][]{williams2013} and in line with deviations from axisymmetry due to spiral structures \citep{faure2014}.

\bsp 

\label{lastpage}

\end{document}